\documentclass[preprint]{revtex4}
\usepackage[latin9]{inputenc}
\usepackage{color}
\usepackage{amsmath}
\usepackage{amssymb}
\usepackage{graphicx}
\usepackage{esint}
\usepackage[unicode=true,pdfusetitle,
 bookmarks=true,bookmarksnumbered=false,bookmarksopen=false,
 breaklinks=false,pdfborder={0 0 1},backref=false,colorlinks=false]
 {hyperref}

\makeatletter
\@ifundefined{textcolor}{}
{%
 \definecolor{BLACK}{gray}{0}
 \definecolor{WHITE}{gray}{1}
 \definecolor{RED}{rgb}{1,0,0}
 \definecolor{GREEN}{rgb}{0,1,0}
 \definecolor{BLUE}{rgb}{0,0,1}
 \definecolor{CYAN}{cmyk}{1,0,0,0}
 \definecolor{MAGENTA}{cmyk}{0,1,0,0}
 \definecolor{YELLOW}{cmyk}{0,0,1,0}
}

\makeatother

\begin{document}

\title{L\'{e}vy flights in the presence of a point sink of finite strength}

\author{Deepika Janakiraman}

\address{ National Centre for Biological Sciences, Bangalore, India}

\begin{abstract}
{In this paper, the absorption of a particle undergoing L\'{e}vy flight in the presence of a point sink of arbitrary strength and position is studied. The motion of  such a particle is given by a modified Fokker-Planck equation whose exact solution in the Laplace domain can be described in terms of the Laplace transform of the unperturbed (absence of the sink) Green's function. This solution for the Green's function is a well-studied, generic result which applies to both fractional and usual Fokker-Planck equations alike. Using this result, the propagator and the absorption time distribution are obtained for free L\'{e}vy flight and L\'{e}vy flight in linear and harmonic potentials in the presence of a delta function sink, and their dependence on the sink strength is analyzed. Analytical results are presented for the long-time behaviour of the absorption time distribution in all the three above mentioned potentials. Simulation results are found to corroborate closely with the analytical results.}
\end{abstract}
\maketitle

\section{Introduction}

Diffusion in the presence of a point sink has always been a problem of great interest as it allows for the evaluation of important quantities like the survival probability and the first-passage time distribution. Herein, the diffusing particle is absorbed (with a certain probability) when it arrives at the sink. Brownian motion in the presence of a point sink is a very well-studied, classic problem \cite{Redner} and has often been used to model chemical reactions in solution which are triggered by a first-passage event, in particular, the problem of electronic relaxation in solution \cite{Sebastian_IASc,Sumi_marcus,Bagchi,Sebastian_CPL}. The reaction-diffusion equation for this problem is given by
\begin{equation}
 \label{Eq:brownian_smoluchowski}
 \frac{\partial P(x,t)}{\partial t}=\left[D\frac{\partial^2}{\partial x^2}+\frac{\partial}{\partial x}\frac{V'(x)}{m\gamma}-k_0\delta(x-x_s)\right]P(x,t),
\end{equation}
where $V(x)$ is the potential, $k_0$ is the sink strength (with dimensions $LT^{-1}$), $x_s$ is the position of the sink, $D$ is the diffusion constant, $\gamma$ is the friction constant, and $m$ is the mass of the particle. The above reaction-diffusion equation is in the overdamped limit including the effect of a sink. One should note that the solution of Eq. (\ref{Eq:brownian_smoluchowski}), $P(x,t)$, is not a normalized quantity as the particle number is not conserved due to absorption.

The probability of absorption of the particle on arrival at the sink is determined by $k_0$. When $k_0\rightarrow \infty$, the diffusing particle is completely annihilated when it reaches $x_s$ and its solution is widely given using the method of images \cite{Risken,Redner} which suggests that
\begin{equation}
 \label{Eq:brownian_method_of_images}
 P(x,t)=G_0(x,t|x_0)-G_0(x,t|x_0-2 x_s),
\end{equation}
where $G_0(x,t|x_0)$ is the propagator in the absence of a sink. This solution results in (i) the density to be exactly equal to zero at $x_s$ and (ii) provides the correct $P(x,t)$ for $(x-x_s)sgn(x_0-x_s)\geq0$, i.e. for all final positions on the same side of the sink as the initial position. For $(x-x_s)sgn(x_0-x_s)<0$, the method of images does not hold and the correct solution in this regime, $P(x,t)=0$ in the case of Brownian motion, is obtained from physical reasoning.

For a sink of finite strength, $k_0$, the solution from the method of images can no longer be applied. This problem was studied in detail by \cite{Sebastian_IASc,Sebastian_CPL} and it was found that the Laplace transform of the Green's function corresponding to the motion in the presence of a finite sink can be given in terms of the Laplace transform of the Green's function in the absence of a sink as
\begin{equation}
 \label{Eq:Greens_function_laplace}
 \mathcal{G}(x,s|x_0)=\mathcal{G}_0(x,s|x_0)-\frac{k_0 \mathcal{G}_0(x,s|x_s)\mathcal{G}_0(x_s,s|x_0)}{1+k_0\mathcal{G}_0(x_s,s|x_s)}.
\end{equation}
The derivation for the above expression, as given in \cite{Sebastian_IASc,Sebastian_CPL},  is provided in Appendix \ref{App:greens_func}. The importance of this result stems from (i) it is an exact result in the Laplace doamain, (ii) provides a generalized solution for diffusive motion in the presence of a delta function sink of arbitrary strength ($k_0$) and position ($x_s$), and (iii) in the time domain, it gives the correct solution for $G(x,t|x_0)$ for all values of $x$, unlike the method of images. Though this result was derived in the context of Brownian motion, it can be applied to any diffusive motion that is Markovian.

The aim of this paper is to study L\'{e}vy flights in the presence of a delta function sink. L\'{e}vy flights is a class of anomalous diffusion where the diffusing particle takes short steps at each interval of time interspersed by occasional, very long \textit{flights}. These step lengths are taken from a L\'{e}vy distribution. A symmetric L\'{e}vy distribution \cite{physrepmet,deepika_Levy_prop,chechkinintroduction2008} is given by
\begin{equation}
 \label{Eq:levy_distribution}
 L_{\alpha,0}(x)=\frac{1}{2\pi}\int^{\infty}_{-\infty}dp\;e^{-|p|^{\alpha}} e^{ipx},
\end{equation}
where $0<\alpha<2$ is the L\'{e}vy index.  When $\alpha=2$, the distribution reduces to a Gaussian which is the step length distribution for Brownian motion. L\'{e}vy distributions are characterized by their fat tails and the asymptotic behaviour of $1/|x|^{\alpha+1}$, resulting in a diverging second moment as opposed to a finite variance for a Gaussian. L\'{e}vy motion can be described by a generalization of the Fokker-Planck equation, referred to as the fractional Fokker-Planck equation \cite{physrepmet,MBK_PRE,deepika_Levy_prop} which is
\begin{equation}
 \label{Eq:Levy_smoluchowski}
 \frac{\partial P(x,t)}{\partial t}=\left\{-D\left(-\frac{\partial^2}{\partial x^2}\right)^{\alpha/2}+\frac{\partial}{\partial x}\frac{V'(x)}{m\gamma}\right\}P(x,t).
\end{equation}
The fractional derivative operator accounts for long flights in the particle trajectory.

The effect of a point sink on L\'{e}vy flights can be captured using a modified Fractional Fokker-Planck equation as it was done for Brownian motion and is given by
\begin{equation}
 \label{Eq:Levy_smoluchowski_sink}
 \frac{\partial P(x,t)}{\partial t}=\left\{-D\left(-\frac{\partial^2}{\partial x^2}\right)^{\alpha/2}+\frac{\partial}{\partial x}\frac{V'(x)}{m\gamma}-k_0\delta(x-x_s)\right\}P(x,t).
\end{equation}
Due to the non-local jumps \cite{leapover_paper}, a L\'{e}vy particle can leap over the sink without actually visiting it. This is fundamentally different from a Brownian particle which, due to its finite variance, necessarily visits the sink before crossing it. The point sink can therefore play two distinct roles on L\'{e}vy motion:

Case 1: \textit{The sink serves as a perfectly absorptive wall} and the particle does not cross $x_s$. This is the problem of L\'{e}vy flights on a semi-infinite line \cite{leapover_paper,levy_half_line} for which the long-time behaviour of the first-passage time distribution (FPTD) is given by the Sparre Andersen theorem \cite{Sparre_Andersen_paper1,Sparre_Andersen_paper2,Redner}. According to the theorem, the asymptotic behaviour of FPTD for any Markov process where the step sizes are chosen from an arbitrary continuous, symmetric distribution behaves as $\sim t^{-3/2}$ which is similar to that for Brownian motion. This theorem goes to suggest  that FPTD shows a universal long-time behaviour, completely independent of the L\'{e}vy index; its coefficient however, depends on $\alpha$ \cite{leapover_paper}.

Case 2: \textit{The point sink absorbs the particle only when it arrives exactly at $x_s$, allowing for leaps over the sink .} As a result of the non-local jumps, passage across the sink does not imply arrival at the sink. In such a situation, one has to be concerned about the \textit{first-arrival time distribution} (FATD) and not FPTD. (It must be noted that FPTD and FATD are identical for Brownian motion) The Green's function for this problem is zero only at the sink and non-zero on its either sides even if the sink strength is infinite. A naive approach to finding the propagator for this problem would be to use the method of images. However, as discussed for Brownian motion, this would give incorrect results for $(x-x_s)sgn(x_0-x_s)<0$. The long-time behaviour of FATD for free L\'{e}vy flight with a perfectly absorbing sink was found to be \cite{metzler_search}
\begin{equation}
 \label{Eq:levy_free_sink}
 p^{free}_{fa}(t)\sim C(\alpha)\frac{|x_0|^{\alpha-1}}{D^{1-\frac{1}{\alpha}} t^{2-\frac{1}{\alpha}}},
\end{equation}
where
\begin{equation}
C(\alpha)=\frac{\alpha\Gamma(2-\alpha)\Gamma\left(2-\frac{1}{\alpha}\right)\sin\left(\frac{\pi\alpha}{2}\right)\sin^2\left(\frac{\pi}{\alpha}\right)}{\pi^2(\alpha-1)}.
\end{equation}
Further, it was found that the solution obtained from the method of images has a $\sim t^{-1-\frac{1}{\alpha}}$ which is neither consistent with the above FATD nor with the Sparre-Anderson theorem.

In this paper, the effect of a point sink on L\'{e}vy flights as described in \textit{Case 2}, is studied. This work goes beyond the finding in Eq. (\ref{Eq:levy_free_sink}) to include delta function sinks of finite strength ($k_0$) at arbitrary locations ($x_s\neq 0$). For this purpose, the prescription for the Green's function given in Eq. (\ref{Eq:Greens_function_laplace}) is used. In the presence of a sink of finite strength, the probability of absorption of the particle upon its first-arrival at the sink is lesser than unity (see  Eq. \ref{Eq:prob_abs}) and therefore, the quantity of interest will be the \textit{absorption time distribution} and not FATD. For a sink of infinite strength, the absorption time distribution will reduce to FATD.

Several situations of physical interest can be studied using fractional diffusion in the presence of a finite sink. One such example is polymer surface diffusion. Single-molecule fluorescence experiments suggest that if the surface contains only a limited number of adsorption sites, the polymers are not completely adsorbed at all times and the diffusion does not occur by polymer-crawling on the surface. The diffusion proceeds via a repeated adsorption-desorption mechanism mediated by diffusion into the bulk \cite{Schwartz1,Schwartz2}. Due to this mechanism, the trajectories of polymer segments on the surface show long jumps which are modeled using a truncated L\'{e}vy flight with $\alpha\approx 1.5$ for the case studied in \cite{Granick}. It was also found that the adsorption sites are not identical; some sites allow for desorption more readily than the others \cite{Shaughnessy} and these `weakly' adsorbing sites may be modeled using finite sinks. Though the problem dealt with in this paper has only one finite point sink, it can be easily extended to an array of finite sinks (like it is in the polymer surface diffusion problem).

Protein diffusion on DNA offers yet another interesting application. DNA being a polymer, has its loop-size distribution given by a fat-tailed L\'{e}vy distribution \cite{Sokolov}. An enzyme such as the DNA polymerase diffuses on the DNA in search of a particular to codon and binds to this site triggering the unzipping process. Due to the presence of loops, the protein could diffuse across to the neck of the loop, referred to as the intersegmental transfer, and the diffusion when viewed from the topology of the DNA, appears to be a L\'{e}vy flight \cite{Lomholtpolymer}. The binding to the target site can be modeled by the delta function sink and the error induced in biological processes due to improper binding can be given by a finite $k_0$.

It is well known that encounters in biology such as foraging, predator-prey dynamics and pollination have significant advantages by adopting L\'{e}vy strategies for their search \cite{GMvish,GMVish_book,Reynolds1}, and they can be studied using a fractional reaction-diffusion equation \cite{metzler_search}. With a finite sink as given in Eq. (\ref{Eq:Levy_smoluchowski_sink}), one can include `non-reactive' encounters, i.e. those encounters which do not halt the diffusive search. Several other examples such as dynamics of infection-transfer, motion of the eye microsaccade which is essential for visual fixation \cite{GMVish_book}, foraging activity by the human brain and memory retrieval \cite{ferraro_PLOS}, and activity in stock markets resulting after the stocks hit a threshold value for the first time \cite{Redner,metzler_search} are potential applications of the problem discussed in this paper.

In section \ref{Sec:sink_diff_pot} of this paper, the role of an absorbing sink on the propagator and absorption time distribution is studied for (i) free particle, (ii) particle subjected to linear potential, and (iii) harmonic potential. The absorption time distribution is given by the negative derivative of the survival probability as
\begin{equation}
 \label{Eq:first_arrival_sur}
 p_{abs}(t)=-\frac{d}{dt}\int^{\infty}_{-\infty}dx\;P(x,t).
\end{equation}
Starting from the propagator given in Eq. (\ref{Eq:Greens_function_laplace}), one can express the absorption time distribution in the Laplace domain as
\begin{equation}
 \label{Eq:first_arrival_derv}
 \tilde{p}_{abs}(s)=\frac{k_0\mathcal{G}_0(x_s,s|x_0)}{1+k_0\mathcal{G}_0(x_s,s|x_s)}=\frac{\mathcal{G}_0(x_s,s|x_0)}{\mathcal{G}_0(x_s,s|x_s)}\frac{1}{\left(1+\frac{1}{k_0\mathcal{G}_0(x_s,s|x_s)}\right)}.
\end{equation}
Given that free L\'{e}vy flight and L\'{e}vy flight in linear and harmonic potentials are exactly solvable problems in the time-domain, we use the knowledge to obtain their Laplace transforms and use them in Eq. (\ref{Eq:Greens_function_laplace}) and Eq. (\ref{Eq:first_arrival_derv}) to evaluate the corresponding quantities. In Section \ref{Sec:simulations}, Langevin dynamics simulations are employed for L\'{e}vy flights  in different potentials and different sink strengths to obtain the corresponding absorption time distributions. The simulation results are found to be in close agreement with the results obtained from (the inverse Laplace transform of) Eq. (\ref{Eq:first_arrival_derv}).

\section{Effect of an absorbing sink on L\'{e}vy flights in different potentials}
\label{Sec:sink_diff_pot}
\subsection{Free L\'{e}vy flight}
The propagator for free L\'{e}vy flight, i.e. $V(x)=0$, can be obtained by solving the fractional Fokker-Planck equation in Eq. \ref{Eq:Levy_smoluchowski} \cite{Fogedby} or using path integrals \cite{deepika_Levy_prop} to obtain
\begin{eqnarray}
 \label{Eq:free_levy_flight}
 G_0^{free}(x,t|x_0)&=&\frac{1}{2\pi}\int^{\infty}_{-\infty}dp\;e^{-D t|p|^{\alpha}} e^{ip(x-x_0)}\nonumber\\
 &=&\frac{1}{(Dt)^{1/\alpha}}L_{\alpha,0}\left(\frac{x-x_0}{(Dt)^{1/\alpha}}\right).
\end{eqnarray}
Alternately, the above L\'{e}vy distribution can be written in terms of a Fox H-function \cite{Fogedby} as
\begin{equation}
 \label{Eq:free_levy_flight_fox_H}
 G_0^{free}(x,t|x_0)=\frac{1}{\alpha|x-x_0|}H^{1,1}_{2,2}\left[\frac{|x-x_0|}{(Dt)^{1/\alpha}}\left|\begin{matrix}
(1,\frac{1}{\alpha})\;\;(1,\frac{1}{2}) \\
(1,1)\;\;(1,\frac{1}{2})
                                                                                                                    \end{matrix}
\right.\right].
\end{equation}
The advantage of using the H-function representation is that its Laplace transform can be represented in terms of yet another H-function \cite{Mathai} (see Appendix \ref{appendix_1}). Using the formula prescribed in Eq. (\ref{Eq:h_function_laplace}), the Laplace transform of the free particle propagator can be easily obtained as
\begin{eqnarray}
 \label{Eq:laplace_transform_free}
 \mathcal{G}_0^{free}(x,s|x_0)&=&L\left\{G_0^{free}(x,t|x_0);s\right\}\nonumber\\
 &=&\frac{s^{-1}}{\alpha|x-x_0|}H^{1,2}_{3,2}\left[\frac{D^{1/\alpha}s^{-1/\alpha}}{|x-x_0|}\left|\begin{matrix}
(0,\frac{1}{\alpha}),(0,1),(0,\frac{1}{2}) \\
(0,\frac{1}{\alpha}),(0,\frac{1}{2})
                                                                                                                    \end{matrix}
\right.\right].
\end{eqnarray}
Further, the Laplace transform of the loop propagator with $x=x_0$ is
\begin{eqnarray}
 \label{Eq:laplace_transform_free_loop_1}
 \mathcal{G}_0^{free}(x,s|x)&=&\frac{1}{2\pi}\int^{\infty}_{-\infty}dp\;\int^{\infty}_{0}dt\;e^{-st}\;e^{-D t|p|^{\alpha}}=\frac{1}{\pi}\int^{\infty}_0dp\;\frac{1}{Dp^{\alpha}+s}\\
  \label{Eq:laplace_transform_free_loop}
 &=&\frac{\csc(\pi/\alpha)}{\alpha D^{\frac{1}{\alpha}}s^{1-\frac{1}{\alpha}}}.
\end{eqnarray}
The Laplace integral for the loop propagator is convergent only for $1<\alpha\leq 2$ as it can be seen from Eq. (\ref{Eq:laplace_transform_free_loop_1}). $\mathcal{G}^{free}_0(x,s|x_0)$, however, is convergent for all values of $\alpha$, viz. $0<\alpha\leq 2$.

\begin{figure}
\includegraphics{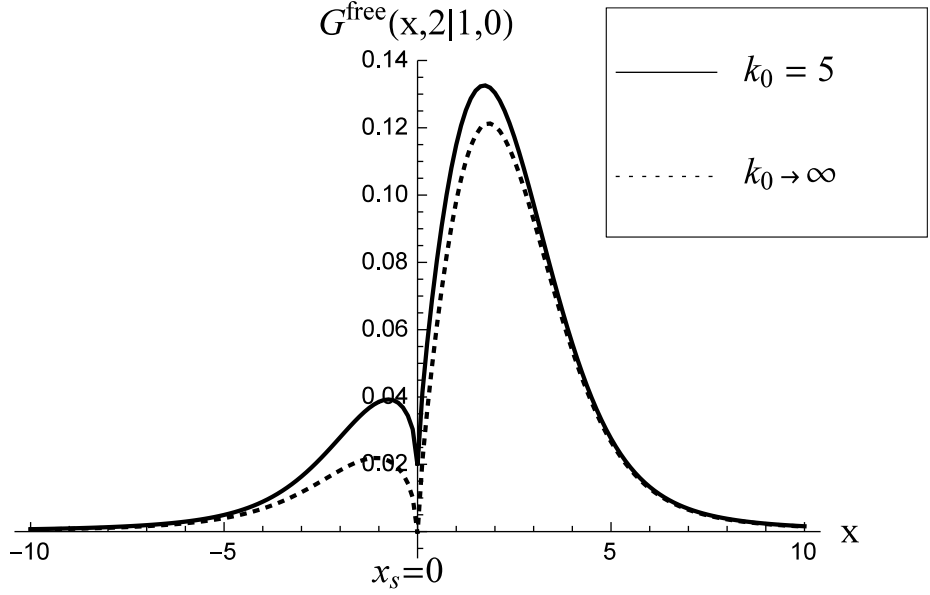}
 \caption{Green's function for free L\'{e}vy flight $\left(\alpha=\frac{5}{3}\right)$ evaluated with the sink at $x_s=0$, $x_0=1$ and time $t=2$ for varying sink strengths. The propagator is zero at the sink for infinite sink strength and non-zero for $k_0=5$.}
\label{Fig:propagator_free_levy}
\end{figure}

Employing Eq. (\ref{Eq:laplace_transform_free}) and Eq. (\ref{Eq:laplace_transform_free_loop}) in Eq. (\ref{Eq:Greens_function_laplace}), one can find the Laplace transform of the propagator for free L\'{e}vy flight in the presence of a point sink. Its inverse Laplace transform has to be obtained to get the desired propagator in the time domain. The inversion, however, cannot be performed analytically. Therefore, the inversion is done numerically using the Gaver-Stehfast (GS) method \cite{Gaver_stehfast} and the propagator is plotted in Fig. \ref{Fig:propagator_free_levy}. As one would expect, the function goes to zero at the sink of infinite strength and for a sink of finite strength, it has a non-zero density at the sink. The propagator is non-zero for $(x-x_s)sgn(x_0-x_s)<0$ (even for the infinite sink case) due to the non-local jumps which are allowed in L\'{e}vy flights.

\begin{figure}
\includegraphics{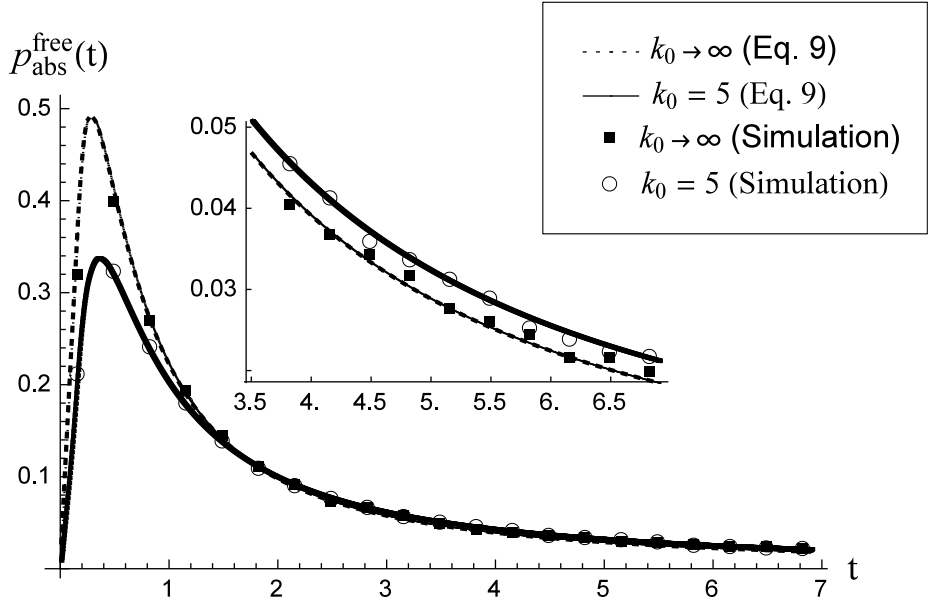}
\caption{Absorption time-distribution for free L\'{e}vy flight $\left(\alpha=\frac{5}{3}\right)$ with $x_s=0$ and $x_0=1$. The tail of $k_0\rightarrow\infty$ resides below that for $k_0=5$. The simulation results are in close agreement with numerically evaluated $p_{abs}^{free}(t)$ using Eq. (\ref{Eq:first_arrival_derv}).}
\label{Fig:first_arrival_free}
\end{figure}

The absorption time distribution for free L\'{e}vy flight in the Laplace domain can be obtained using Eq. (\ref{Eq:first_arrival_derv}) which is
\begin{equation}
\label{Eq:first_arrival_free_levy}
\tilde{p}^{free}_{abs}(s)=\frac{D^{1/\alpha}\;s^{-1/\alpha}\;\sin(\pi/\alpha)}{|x_s-x_0|\left(1+\frac{\alpha D^{\frac{1}{\alpha}}s^{1-\frac{1}{\alpha}}}{k_0\;\csc(\pi/\alpha)}\right)}\;H^{1,2}_{3,2}\left[\frac{D^{1/\alpha}s^{-1/\alpha}}{|x_s-x_0|}\left|\begin{matrix}
\left(0,\frac{1}{\alpha}\right),(0,1),\left(0,\frac{1}{2}\right) \\
\left(0,\frac{1}{\alpha}\right),\left(0,\frac{1}{2}\right)
                                                                                                                    \end{matrix}
\right.\right].
\end{equation}
Though the above Laplace transform cannot be inverted analytically, one can obtain the long-time behaviour of the absorption time distribution from its small-$s$ limit (see Eq. (\ref{Eq:appendix_free_laplace}) for details) which is given by
\begin{equation}
\label{Eq:first_arrival_free_levy_1}
\tilde{p}^{free}_{abs}(s)=1+\left(\mathcal{A}_1^{free}+\frac{\mathcal{A}_2^{free}}{k_0}\right)s^{1-\frac{1}{\alpha}}+..,
\end{equation}
where
\begin{equation}
\label{Eq:free_levy_laplace_coefficients}
\mathcal{A}^{free}_1=\frac{\alpha\Gamma\left(1-\alpha\right)\sin\left(\frac{\pi}{\alpha}\right)}{\Gamma\left(1-\frac{\alpha}{2}\right)\Gamma\left(\frac{\alpha}{2}\right)}\;\frac{|x_s-x_0|^{\alpha-1}}{D^{1-\frac{1}{\alpha}}},\;\;\;\;\;\mathcal{A}_2^{free}=-\alpha D^{\frac{1}{\alpha}}\sin\left(\frac{\pi}{\alpha}\right).
\end{equation}
The Laplace inverse of the above equation (see Eq. (\ref{Eq:appendix_free_time})) which gives the long-time behaviour of $p_{abs}^{free}(t)$ turns out to be
\begin{equation}
 \label{Eq:first_arrival_long_time}
 p^{free}_{abs}(t)\sim\left(\mathcal{C}^{free}_1+\frac{\mathcal{C}^{free}_2}{k_0}\right)\frac{1}{t^{2-1/\alpha}},
\end{equation}
where
\begin{eqnarray}
\label{Eq:free_levy_coefficients}
\mathcal{C}_1^{free}&=&\frac{\alpha\Gamma\left(2-\alpha\right)\Gamma\left(2-\frac{1}{\alpha}\right)\sin\left(\frac{\pi\alpha}{2}\right)\sin^2\left(\frac{\pi}{\alpha}\right)}{\pi^2\left(\alpha-1\right)}\;\frac{|x_s-x_0|^{\alpha-1}}{D^{1-\frac{1}{\alpha}}}\\
&=&\frac{(\alpha-1)|x_s-x_0|^{\alpha-1}}{D^{1-\frac{1}{\alpha}}}\left(\frac{\pi\;\Gamma(3-\alpha)}{2\alpha\;\Gamma\left(2-\frac{1}{\alpha}\right)\Gamma\left(2-\frac{\alpha}{2}\right)\Gamma^2\left(\frac{1}{\alpha}\right)\Gamma\left(\frac{\alpha}{2}\right)}\right)\\
\mathcal{C}_2^{free}&=&(\alpha-1)^2\left(\frac{\pi D^{\frac{1}{\alpha}}}{\alpha\Gamma\left(2-\frac{1}{\alpha}\right)\Gamma^2\left(\frac{1}{\alpha}\right)}\right)
\end{eqnarray}
This is a general result for any sink strength, $k_0$, and any sink position, $x_s$. The first term of this result is the asymptotic behaviour of $p_{abs}^{free}(t)$ in the presence of a sink of infinite strength and reduces to the result in Eq. (\ref{Eq:levy_free_sink}) when $x_s=0$. The second term is a correction to the long-time behaviour in the presence of a sink of finite strength. For $1<\alpha\leq2$, this term is always positive and suggests that smaller the sink strength, larger is the value of $p^{free}_{abs}(t)$ at long times. Fig. \ref {Fig:first_arrival_free} illustrates this finding. The plot contains the absorption time distribution for $k_0=5$ and $k_0\rightarrow\infty$  evaluated numerically by inverting Eq. (\ref{Eq:first_arrival_free_levy}) using the GS method. $p^{free}_{abs}(t)$ curve for $k_0\rightarrow\infty$ (dotted line) lies below that for $k_0=5$ curve (solid line) at long times, in agreement with our expectation. One must note that at long times the absorption time distribution has a $\sim t^{-2+1/\alpha}$ behaviour. The power-law decay depends solely on the $\alpha$  value irrespective of the sink strength. For $0<\alpha\leq 1$, $p_{abs}^{free}(s)$ goes to zero for all $\alpha$ as the denominator, $\mathcal{G}_0^{free}(x,s|x)$, diverges. It must be interpreted as a free particle undergoing L\'{e}vy flight with $0<\alpha\leq 1$ will never find a point target and its motion will be oblivious to the presence of a point sink \cite{search_levy_brown}.

\subsection{L\'{e}vy flight in a linear potential}
\label{Subsec:lin_pot}

The fractional Fokker-Planck equation (or the path integral) for L\'{e}vy flight in a linear potential, $V(x)=-F x$, can be exactly solved to obtain the propagator \cite{deepika_Levy_prop,Fogedby} as
\begin{eqnarray}
 \label{Eq:Linear_potential_propagator}
 G^{lin}_0(x,t|x_0,0)&=&\frac{1}{2\pi}\int^{\infty}_{-\infty}dp\;e^{-D t|p|^{\alpha}} e^{i p \left(x-x_0-F t/\gamma\right)}\nonumber\\
&=&\frac{1}{(Dt)^{1/\alpha}}\;L_{\alpha,0}\left(\frac{x-x_0-F t/\gamma}{(Dt)^{1/\alpha}}\right).
\end{eqnarray}
(Mass of the particle is taken to be unity.) Unlike the free L\'{e}vy flight problem, the Laplace transform of the above propagator ($\mathcal{G}_0^{lin}(x,s|x_0)$) cannot be written as a closed form Fox H-fuction for arbitrary values of $x$ and $x_0$, although it can be evaluated numerically. The Laplace transform of the loop propagator with $x=x_0$ can be expressed as a Fox H-function and evaluated up to arbitrary precision.
\begin{eqnarray}
\label{Eq:loop_propagator_linear_laplace}
\mathcal{G}^{lin}_0(x,s|x)&=& L\left\{G_0^{lin}(x,s|x);s\right\}\nonumber\\
&=&\int^{\infty}_0 dt\;e^{-st}\;\frac{\gamma t^{-1}}{\alpha |F|}\;H^{1,1}_{2,2}\left[\frac{|F| t}{\gamma(Dt)^{1/\alpha}}\left|\begin{matrix}
(1,\frac{1}{\alpha})\;\;(1,\frac{1}{2}) \\
(1,1)\;\;(1,\frac{1}{2})
                                                                                                                    \end{matrix}
\right.\right]\nonumber\\
&=&\frac{\gamma}{\alpha|F|}H^{1,2}_{3,2}\left[\frac{|F|}{\gamma}\frac{s^{-1+1/\alpha}}{ D^{1/\alpha}}\left|\begin{matrix}
(1,1-\frac{1}{\alpha}),(1,\frac{1}{\alpha}),(1,\frac{1}{2}) \\
(1,1),(1,\frac{1}{2})
                                                                                                                    \end{matrix}
\right.\right].
\end{eqnarray}

\begin{figure}
\includegraphics{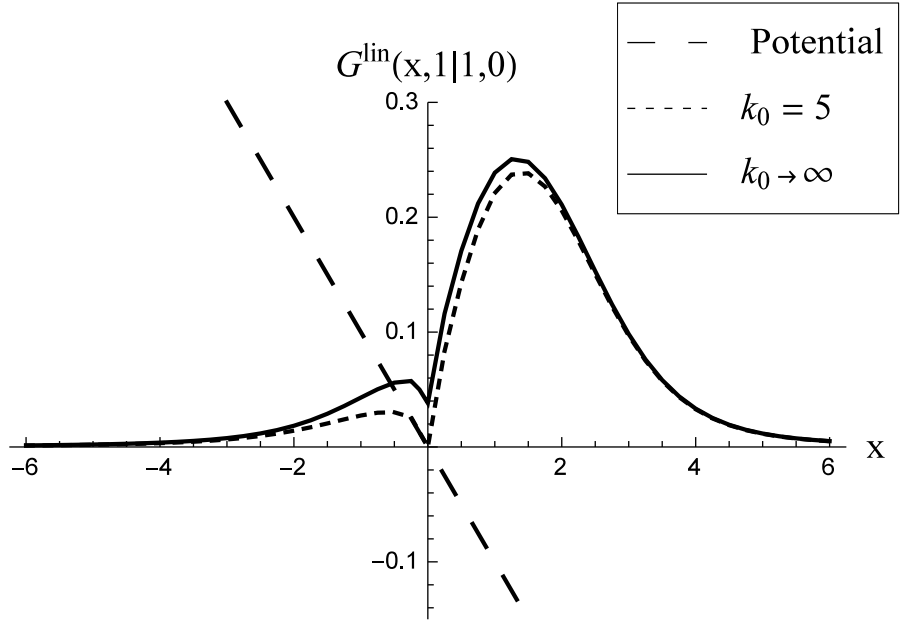}
\caption{Propagator for L\'{e}vy flight $\left(\alpha=\frac{5}{3}\right)$ in a linear potential with $F=1.0$ evaluated with the sink at $x_s=0$, $x_0=1$ and time $t=1$ for varying sink strengths. The propagator is zero at the sink for infinite sink strength and non-zero for $k_0=5$.}
\label{Fig:propagator_force}
\end{figure}

On employing the appropriate expressions of Laplace transforms in Eq. (\ref{Eq:Greens_function_laplace}) and inverting it numerically, the  propagator is obtained in the time domain and plotted in Fig. \ref{Fig:propagator_force}. As observed with free L\'{e}vy flight, the probability density goes to zero at the sink when the strength is infinity and is non-zero for a sink of finite strength.

\begin{figure}
\includegraphics{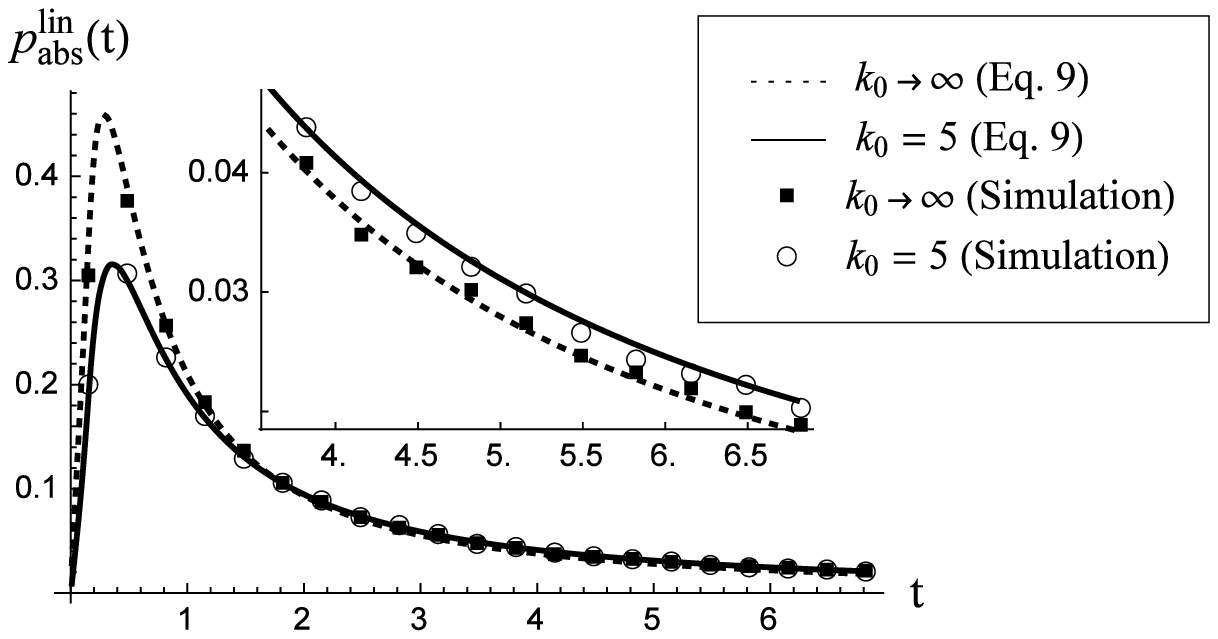}
 \caption{Absorption time distribution for L\'{e}vy flight $\left(\alpha=\frac{5}{3}\right)$ in a linear potential with $F=1.0$ evaluated with $x_s=0$ and $x_0=1$. The tail of $k_0\rightarrow\infty$ resides below that for $k_0=5$. The simulation results are in close agreement with numerically evaluated $p_{abs}^{lin}(t)$ using Eq. (\ref{Eq:first_arrival_derv}).}
\label{Fig:first_arrival_force}
\end{figure}

\begin{figure}
\includegraphics{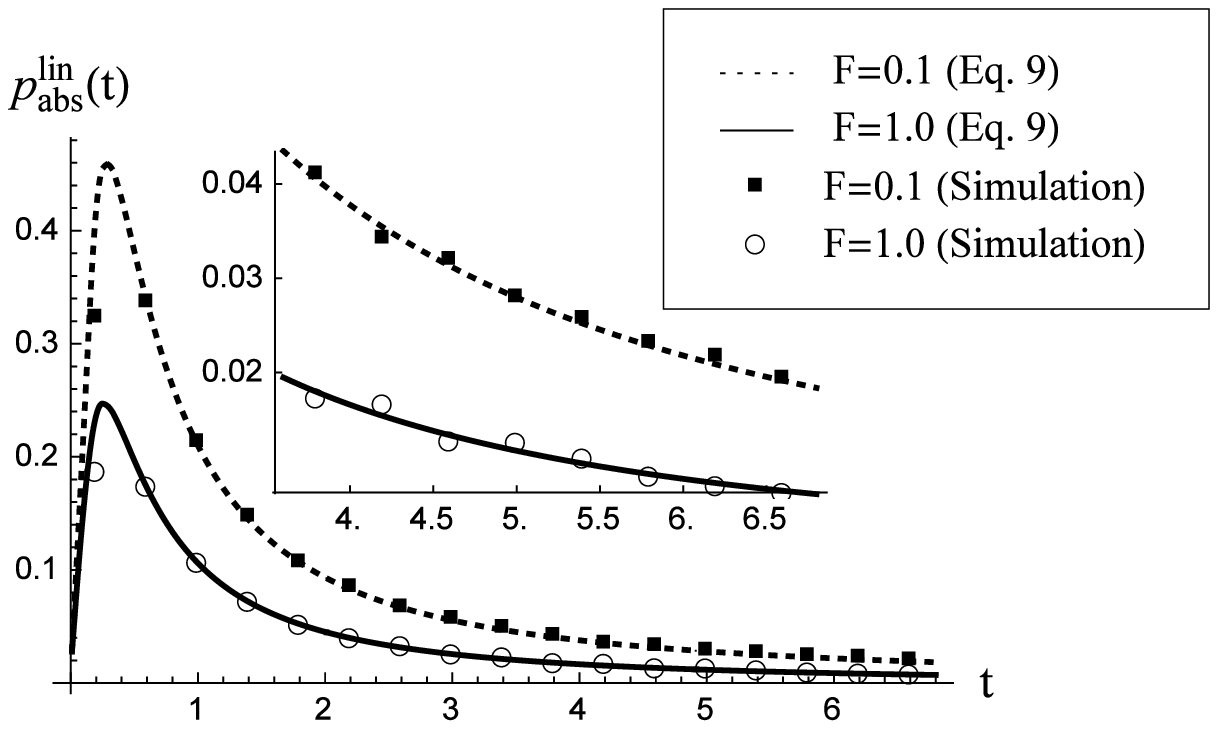}
\caption{Absorption time distribution for L\'{e}vy flight $\left(\alpha=\frac{5}{3}\right)$ in a linear potential with different slopes evaluated with $x_s=0$, $x_0=1$ and $k_0\rightarrow\infty$. The tail of $F=1.0$ resides below that of $F=0.1$. The simulation results are in close agreement with numerically evaluated $p_{abs}^{lin}(t)$ using Eq. (\ref{Eq:first_arrival_derv}).}
\label{Fig:first_arrival_diff_force}
\end{figure}

The absorption time distribution is formally given by
\begin{equation}
\label{Eq:first_arrival_lin_a}
p_{abs}^{lin}(t)=L^{-1}\left\{\frac{k_0\;\mathcal{G}_0^{lin}(x_s,s|x_0)}{1+k_0\;\mathcal{G}_0^{lin}(x_s,s|x_s)};t\right\}.
\end{equation}
Though  $\mathcal{G}_0^{lin}(x_s,s|x_0)$ does not have a closed form expression, the small-$s$ behaviour of $\tilde{p}_{abs}^{lin}(s)$ can be evaluated analytically for small values of the force such that $F<<1$ (or $F^2\approx 0$ that it can be neglected) (see Appendix \ref{appendix_3} for the derivation). The expansion of $\tilde{p}^{lin}_{abs}(s)$ valid for small-$s$ is
\begin{equation}
\label{Eq:first_arrival_lin_1}
\tilde{p}^{lin}_{abs}(s)=1+\left(\mathcal{A}_1^{lin}+\frac{\mathcal{A}_2^{lin}}{k_0}+\frac{F}{\gamma}(x_s-x_0) \mathcal{A}_3^{lin}\right)s^{1-\frac{1}{\alpha}}+\frac{F}{\gamma}(x_s-x_0)\mathcal{A}^{lin}_4\;s^{-1+\frac{2}{\alpha}}+...,
\end{equation}
where
\begin{eqnarray}
\mathcal{A}^{lin}_1&=&\frac{\alpha\Gamma\left(1-\alpha\right)\sin\left(\frac{\pi}{\alpha}\right)}{\Gamma\left(1-\frac{\alpha}{2}\right)\Gamma\left(\frac{\alpha}{2}\right)}\;\frac{|x_s-x_0|^{\alpha-1}}{D^{1-\frac{1}{\alpha}}},\;\;\;\;\;\mathcal{A}_2^{lin}=-\alpha D^{\frac{1}{\alpha}}\sin\left(\frac{\pi}{\alpha}\right),\nonumber\\
\mathcal{A}_3^{lin}&=&\frac{\alpha\Gamma\left(2-2\alpha\right)\sin\left(\frac{\pi}{\alpha}\right)\sin(\pi\alpha)}{\pi}\;\frac{|x_s-x_0|^{2\alpha-3}}{D^{2-1/\alpha}},\;\;\;\;\;\mathcal{A}^{lin}_4=\frac{\Gamma\left(2-\frac{3}{\alpha}\right)\Gamma\left(\frac{3}{\alpha}\right)\sin\left(\frac{\pi}{\alpha}\right)}{\pi D^{\frac{2}{\alpha}}}.\nonumber\\
&&
\end{eqnarray}
The above expression can be easily inverted to obtain the long-time behaviour of $p_{abs}^{lin}(t)$ which is
 \begin{eqnarray}
 \label{Eq:first_arrival_long_time_lin}
 p^{lin}_{fa}(t)&\sim&\left(\mathcal{C}^{lin}_1+\frac{\mathcal{C}^{lin}_2}{k_0}\right)\frac{1}{t^{2-1/
\alpha}}+\frac{F}{\gamma}(x_s-x_0)\left(\frac{\mathcal{C}^{lin}_3}{t^{2-1/\alpha}}-\frac{\mathcal{B}^{lin}}{t^{\frac{2}{\alpha}}}\right),
\end{eqnarray}
where
\begin{eqnarray}
\label{Eq:lin_levy_coefficients}
\label{C_1_lin}
\mathcal{C}_1^{lin}&=&\frac{(\alpha-1) |x_s-x_0|^{\alpha-1}}{D^{1-\frac{1}{\alpha}}}\left(\frac{\pi\;\Gamma(3-\alpha)}{2\alpha\;\Gamma\left(2-\frac{1}{\alpha}\right)\Gamma\left(2-\frac{\alpha}{2}\right)\Gamma^2\left(\frac{1}{\alpha}\right)\Gamma\left(\frac{\alpha}{2}\right)}\right)\\
\label{C_2_lin}
\mathcal{C}_2^{lin}&=&(\alpha-1)^2\;D^{\frac{1}{\alpha}}\left(\frac{\pi}{\alpha\Gamma\left(2-\frac{1}{\alpha}\right)\Gamma^2\left(\frac{1}{\alpha}\right)}\right)\\
\label{C_3_lin}
\mathcal{C}_3^{lin}&=&\;\frac{(\alpha-1)^2|x_s-x_0|^{2\alpha-3}}{(2\alpha-3)D^{2-\frac{1}{\alpha}}}\left(\frac{\pi\Gamma(5-2\alpha)}{4\alpha\;\Gamma(\alpha)\Gamma(3-\alpha)\Gamma\left(2-\frac{1}{\alpha}\right)\Gamma^2\left(\frac{1}{\alpha}\right)}\right)\\
\label{B_lin}
\mathcal{B}^{lin}&=&\frac{(2-\alpha)(\alpha-1)}{(2\alpha-3)D^{\frac{2}{\alpha}}}\left(\frac{2\;\Gamma\left(4-\frac{3}{\alpha}\right)\Gamma\left(\frac{3}{\alpha}\right)}{3\alpha\;\Gamma\left(3-\frac{2}{\alpha}\right)\Gamma\left(2-\frac{1}{\alpha}\right)\Gamma\left(\frac{1}{\alpha}\right)}\right).
\end{eqnarray}
The coefficients $C^{lin}_1=C^{free}_1$ and $C^{lin}_2=C^{free}_2$, and the asymptotic power-law decay of the force-independent terms are identical to the free L\'{e}vy flight problem in the presence of a point sink .

 For small values of $F$, the presence of the ramp appears as a correction to free L\'{e}vy flight. The correction has two contributions, one which decays as $\sim t^{-2+\frac{1}{\alpha}}$ and the other as $\sim t^{-\frac{2}{\alpha}}$. In order to analyze the consequences of these terms, the absorption time distribution is written as

 \begin{eqnarray}
 \label{Eq:first_arrival_long_time_lin_1}
 p^{lin}_{fa}(t)&\sim&\left(\mathcal{C}^{lin}_1+\frac{\mathcal{C}^{lin}_2}{k_0}\right)\frac{1}{t^{2-1/
\alpha}}+\frac{F}{\gamma}(x_s-x_0)\frac{\mathcal{C}^{lin}_3}{t^{2-1/\alpha}}\left(1-\frac{\mathcal{B}^{lin}}{\mathcal{C}_3^{lin}}t^{2-\frac{3}{\alpha}}\right).
\end{eqnarray}
\begin{enumerate}
\item{\textit{For $\frac{3}{2}<\alpha\leq2$}:  In this range of $\alpha$, the exponent $2/\alpha$ is smaller that $2-1/\alpha$, implying that the former will be dominant in the long-time decay. However, the ratio $\mathcal{B}^{lin} t^{2-\frac{3}{\alpha}}/\mathcal{C}_3^{lin}<<1$ in the vicinity of $\alpha=2$, unless $t$ is extremely large, and vanishes exactly for $\alpha=2$ (see Eq. (\ref{B_lin})). This implies that the origin of $\mathcal{B}^{lin}$ is due to the non-local jumps in a L\'{e}vy flight. For $\alpha$ close to $2$, the decay follows $\sim\mathcal{C}_3^{lin}t^{-2+\frac{1}{\alpha}}$ until very large $t$ before $\sim\mathcal{B}^{lin}t^{-\frac{2}{\alpha}}$ can take over. When $\alpha=2$, the power-law decay of the F-dependent term goes as $\sim t^{-3/2}$, in agreement with the standard results of Brownian motion \cite{bruce_berne}.}
\item{\textit{For $\alpha=\frac{3}{2}$}: The quantity $\mathcal{C}_3^{lin}\left(1-\frac{\mathcal{B}^{lin}}{\mathcal{C}_3^{lin}}t^{2-\frac{3}{\alpha}}\right)$ is well-defined in the limit $\alpha\rightarrow 3/2$ and the decay follows $\sim t^{-4/3}$ for large $t$.  }
\item{\textit{For $1<\alpha<\frac{3}{2}$}: In this regime, the decay of $t^{2-\frac{1}{\alpha}}$ will be slower that that of $t^{-\frac{2}{\alpha}}$. However, in the vicinity of $\alpha=1$, $\mathcal{C}_3^{lin}$ goes to zero quadratically (see Eq. (\ref{C_3_lin})) while $\mathcal{B}^{lin}$ goes to zero linearly (see Eq. (\ref{B_lin})). Therefore, the decay will initially obey $\sim \mathcal{B}^{lin} t^{-\frac{2}{\alpha}}$ power-law and further $\sim \mathcal{C}_3^{lin} t^{-2+\frac{1}{\alpha}}$ will take over at long times.}
\item{\textit{For $0<\alpha\leq 1$}: In the limit of small $F$, the Laplace transform of the loop propagator is divergent in this range of $\alpha$ (see Eq. (\ref{Eq:small_force_laplace_loop})). This implies that $p_{abs}^{lin}(t)=0$ for all values of $t$ when $0<\alpha\leq 1$.}
 \end{enumerate}

\textbf{Consequences on L\'{e}vy based target search:} The asymptotic behaviour in Eq. (\ref{Eq:first_arrival_long_time_lin_1}) can have very interesting effects on L\'{e}vy flights inspired target search methods at long times. The goal is to find an optimal value of $\alpha$ that can add density to the tail of $p_{abs}^{lin}$ making the target search better at long times. Both $\mathcal{C}_3^{lin}$ and $\mathcal{B}^{lin}$ are strictly $+ve$ when $\frac{3}{2}<\alpha\leq 2$ and stricly $-ve$ when $1<\alpha<\frac{3}{2}$.  For a downstream target, $x_s-x_0>0$,  Eq. (\ref{Eq:first_arrival_long_time_lin_1}) suggests that there are two ways in which one can obtain an additive contribution to the tail of $p_{abs}^{lin}(t)$ from the F-dependent terms: (i) employ $\alpha$ close to $2$ such that $\mathcal{C}_3^{lin}$ is positive and $\mathcal{B}^{lin} t^{2-\frac{3}{\alpha}}/\mathcal{C}_3^{lin}<<1$ or (ii) employ $\alpha\approx 1$ such that $\mathcal{C}_3^{lin}$ is negative and  $\mathcal{B}^{lin} t^{2-\frac{3}{\alpha}}/\mathcal{C}_3^{lin}>>1$. In the latter, the contribution to the tail, however $+ve$, will approach zero linearly close to $\alpha=1$ and is therefore, insignificant. Hence, it is beneficial to use the former, $\alpha=2$ - a Brownian search, for a downstream target such that $\mathcal{B}^{lin} t^{2-\frac{3}{\alpha}}/\mathcal{C}_3^{lin}=0$, removing any negative contribution to the tail of the absorption time distribution. For an upstream target, it is apparent from Eq. (\ref{Eq:first_arrival_long_time_lin_1}) that a purely Brownian search will only lead to a negative contribution to the tail from the terms which are F-dependent. A search with $\alpha=1$ will also lead to a vanishingly small $p_{abs}^{lin}(t)$. An optimal value of $\alpha$ for such a search will depend on the distance of separation $(x_s-x_0)$, $F$, $D$ and $t$. A detailed discussion on the optimal value of $\alpha$ for target search on a ramp is provided in \cite{search_levy_brown}. The results were interpreted in terms of a search efficiency parameter given  by
\begin{equation}
\label{Eq:efficiency}
\mathcal{E}=\langle t^{-1}\rangle=\int^{\infty}_{0}ds\;p_{abs}^{lin}(s),
\end{equation}
and the findings in this paper from $p_{abs}^{lin}(t)$ regarding upstream and downstream targets are consistent with those of \cite{search_levy_brown}. (The notion of upstream or downstream is presented keeping $F>0$. If $F<0$, this notion has to be reversed.)

Absorption time distributions obtained from numerical evaluation of Eq. (\ref{Eq:first_arrival_lin_a}) are plotted in Fig. \ref{Fig:first_arrival_force} and Fig. \ref{Fig:first_arrival_diff_force}. In Fig. \ref{Fig:first_arrival_force}, the distributions are obtained for $F=1.0$ and sink strengths $k_0=5$ and $k_0\rightarrow \infty$. The tail of $k_0\rightarrow\infty$ is below that of $k_0=5$ which is similar to the free L\'{e}vy flight problem and can be explained with the same reasoning. In Fig. \ref{Fig:first_arrival_diff_force}, the curves are obtained for $F=1.0$ and $F=0.1$ with a point sink of infinite strength. The sink is placed at the origin and the initial position is $x_0=1$, i.e. the sink is upstream to the initial position and it is found that the tail of $F=1.0$ curve resides below that of $F=0.1$. One can interpret this observation in terms of Eq. (\ref{Eq:first_arrival_long_time_lin}) as follows: Given that $\alpha=5/3$ in the plot, both $\mathcal{C}_3^{lin}$ and $\mathcal{B}^{lin}$ are $+ve$ quantities. Further, $(x_s-x_0)<0$ in the plot. For the range of $t$ plotted in the figure, $F\left(x_s-x_0\right)\mathcal{C}_3^{lin}/(\gamma t^{2-\frac{1}{\alpha}})$ is the dominant term which results in a larger subtractive correction to the tail of $p_{abs}^{lin}(t)$ for a larger $F$, causing the asymptote of $F=1.0$ to lie below that of $F=0.1$.

\subsection{L\'{e}vy flight in a Harmonic potential}

L\'{e}vy flight in a harmonic potential, $V(x)=\lambda x^2/2$, is again a solvable problem for which the propagator is given by
\begin{equation}
 \label{Eq:levy_flights_prop_harmonic}
 G_0^{har}(x,t|x_0)=\left(\frac{\alpha\lambda}{D\gamma\left(1-e^{-\alpha\lambda t}\right)}\right)^{1/\alpha}\;L_{\alpha,0}\left(\frac{x-x_0e^{-\lambda t}}{\left(D\gamma(1-e^{-\alpha\lambda t})/(\alpha\lambda)\right)^{1/\alpha}}\right).
\end{equation}
(The mass is assumed to be unity.) It was identified in \cite{Deepika_sebastian_PRE2014}, that this propagator can be written as a sum over the eigenvalues and eigenfunctions as
\begin{equation}
 \label{Eq:levy_flights_prop_harmonic_sum}
 G_0^{har}(x,t|x_0)=\sum_{n,m=0}^{\infty}\frac{(-x_0/D^{\frac{1}{\alpha}})^n}{\Gamma(n+1)\Gamma(m+1)}\psi_{n,m}(x)e^{-(n+m\alpha)\frac{\lambda t}{\gamma}},
\end{equation}
where
\begin{equation}
 \label{Eq:eigenfunction_LOUP}
 \psi_{n,m}(x)=\frac{1}{2\pi}\int^{\infty}_{-\infty}dp\;e^{-|p|^{\alpha}\gamma/(\alpha\lambda)}\;|p|^{m\alpha}(ip)^ne^{ipx/D^{1/\alpha}}.
\end{equation}
It was interesting to note that this one-dimensional problem has its eigenvalues denoted by two indices $n, m$ and was discussed in detail in \cite{Deepika_sebastian_PRE2014}.

\begin{figure}
\includegraphics{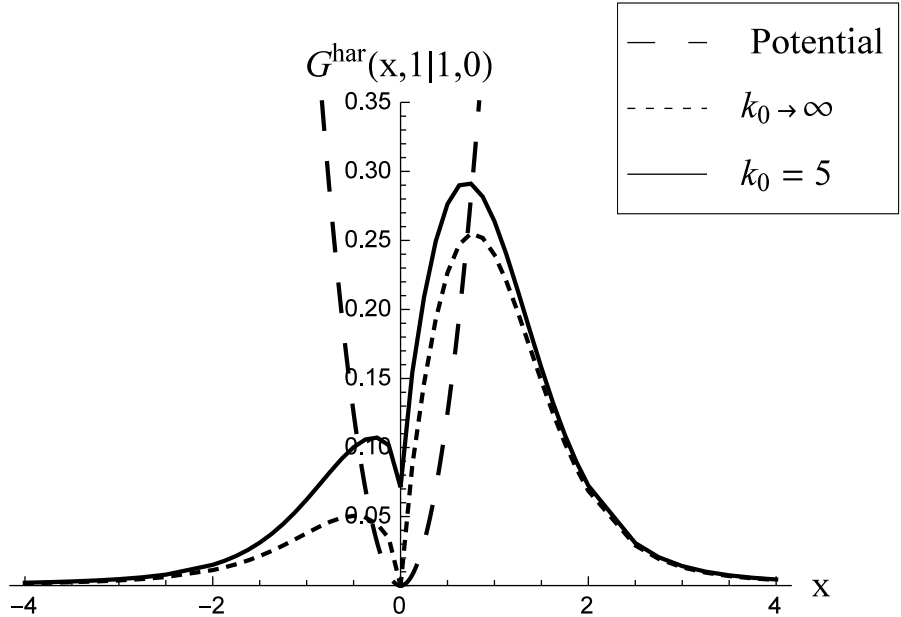}
\caption{Green's function for L\'{e}vy flight $\left(\alpha=\frac{5}{3}\right)$ in a harmonic potential with $\lambda=1.0$ evaluated with the sink at $x_s=0$, $x_0=1$ and time $t=1$ for varying sink strengths. The propagator is zero at the sink for infinite sink strength and non-zero for $k_0=5$.}
\label{Fig:propagator_har}
\end{figure}

Once the propagator is written in the form of Eq. (\ref{Eq:levy_flights_prop_harmonic_sum}), it is trivial to obtain its Laplace transform as
\begin{equation}
 \label{Eq:Levy_harmonic_laplace}
 \mathcal{G}_0^{har}(x,s|x_0)=\sum_{n,m=0}^{\infty}\frac{(-x_0/D^{\frac{1}{\alpha}})^n}{\Gamma(n+1)\Gamma(m+1)}\psi_{n,m}(x)\frac{1}{s+(n+m\alpha)\lambda/\gamma}.
\end{equation}
As for the loop propagator which is necessary for the calculation, it reduces to an elegant closed form expression when the sink is placed a the origin which is given by
\begin{equation}
\label{Eq:Levy_harmonic_loop}
 \mathcal{G}_0^{har}(0,s|0)=\left(\frac{\alpha\lambda}{D\gamma}\right)^{1/\alpha}\frac{\csc(\pi/\alpha)}{s\alpha}
 \frac{\Gamma\left(1+\frac{s\gamma}{\alpha\lambda}\right)}{\Gamma(1-\frac{1}{\alpha}+\frac{s\gamma}{\alpha\lambda})}.
\end{equation}
The above expression is valid only when $1<\alpha\leq 2$. For $0<\alpha\leq 1$, the Laplace integral is divergent (see Appendix \ref{appendix_4}). \textit{Due to the ease of evaluating $\mathcal{G}^{har}_0(0,s|0)$, all the results presented in this section will have the sink at $x_s=0$.}
The propagator is evaluated by inserting Eq. (\ref{Eq:Levy_harmonic_laplace}) and Eq. (\ref{Eq:Levy_harmonic_loop}) in Eq. (\ref{Eq:Greens_function_laplace}), and numerically inverting the Laplace transform. The results are plotted in Fig. \ref{Fig:propagator_har} and as observed in the previous problems, the density at the sink is zero when $k_0\rightarrow\infty$ and assumes a non-zero value for a sink of finite strength.

\begin{figure}
\includegraphics{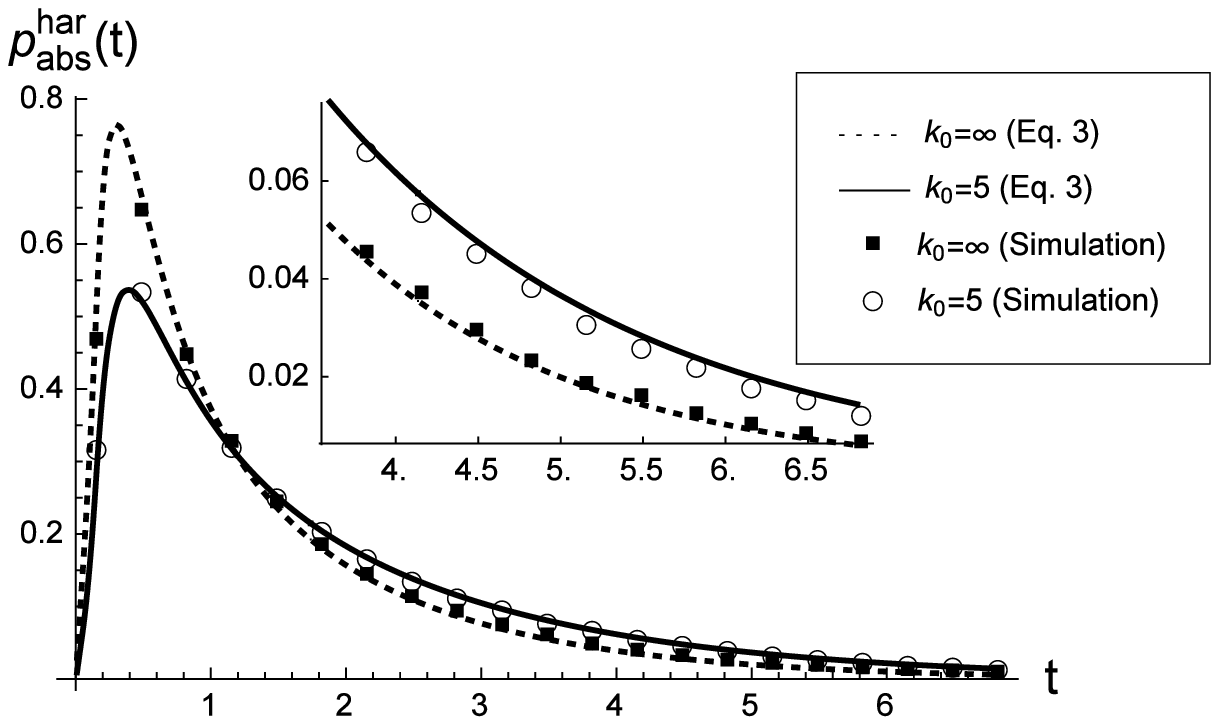}
\caption{Absorption time distribution for L\'{e}vy flight $\left(\alpha=\frac{5}{3}\right)$ in a harmonic potential with $\lambda=1.0$ evaluated with $x_s=0$ and $x_0=1$. The tail of $k_0\rightarrow\infty$ resides below that for $k_0=5$. The simulation results are in close agreement with numerically evaluated $p_{abs}^{har}(t)$ using Eq. (\ref{Eq:first_arrival_derv}).}
\label{Fig:first_arrival_har}
\end{figure}

\begin{figure}
\includegraphics{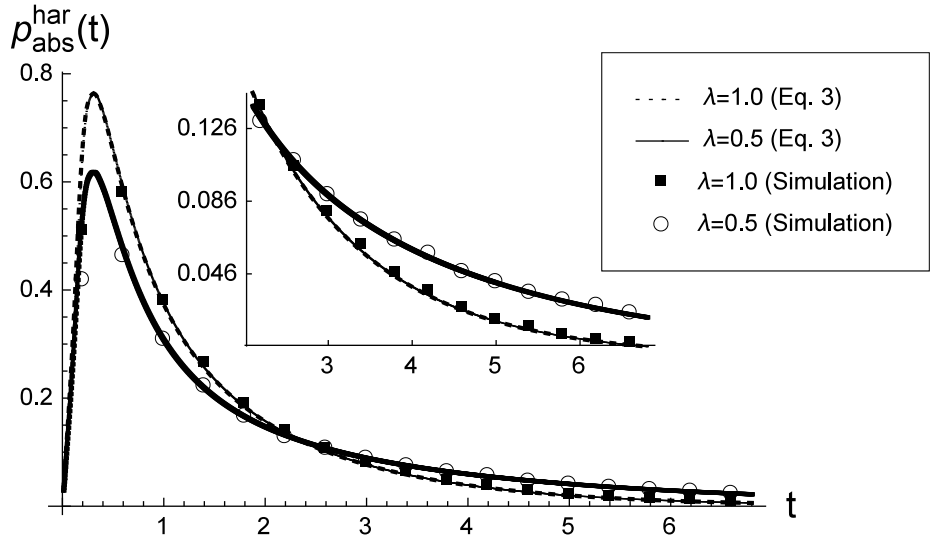}
\caption{Absorption time distribution for L\'{e}vy flight $\left(\alpha=\frac{5}{3}\right)$ in a harmonic potential with different force constants evaluated with $x_s=0$, $x_0=1$ and $k_0\rightarrow\infty$. The tail of $\lambda=1.0$ resides below that of $\lambda=0.5$. The simulation results are in close agreement with numerically evaluated $p_{abs}^{har}(t)$ using Eq. (\ref{Eq:first_arrival_derv}).}
\label{Fig:first_arrival_har_diff}
\end{figure}

The absorption time distribution is evaluated using the appropriate form of Eq. (\ref{Eq:first_arrival_derv}) and numerically evaluating the inverse Laplace transform. The trend for $p_{abs}^{har}(t)$ in the presence of finite and infinite sinks is plotted in  Fig. \ref{Fig:first_arrival_har} and is found to be similar to the previous problems which can be explained with similar arguments. The plot for $p_{abs}^{har}(t)$ in harmonic potentials with different force constants is shown in Fig. \ref{Fig:first_arrival_har_diff}. For the potential with a larger force constant ($\lambda=1.0$ in the plot), the drift to the sink, which is placed at the origin, is larger and therefore finds the sink at shorter times  reducing the density at longer times when compared to the potential with a smaller $\lambda$ ($\lambda=0.5$ in the plot).

The observations from numerical results can be proven analytically. The long-time behaviour of the absorption time distribution is obtained by evaluating the dominant term for large $t$ from the contour integral representing the Laplace inverse which is given by
\begin{equation}
\label{Eq:Levy_harmonic_contour}
p_{abs}^{har}(t)=\frac{1}{2\pi i}\int^{c+i\infty}_{c-i\infty} ds\;e^{st}\frac{k_0 \mathcal{G}_0^{har}(0,s|x_0)}{1+k_0 \mathcal{G}_0^{har}(0,s|0)},
\end{equation}
where the path of integration is a straight line $Re(s)=c$ with c is greater than the real part of all singularities of the integrand.
At large times, the behaviour is (see Appendix \ref{appendix_4} for details)
\begin{equation}
\label{Eq:Levy_harmonic_asymptotic}
p_{abs}^{har}(t)\sim\left(\mathcal{C}^{har}_1+\frac{\mathcal{C}^{har}_2}{k_0}\right) e^{-(\alpha -1) \lambda t/\gamma},
\end{equation}
where
\begin{eqnarray}
\mathcal{C}_1^{har}&=&\alpha\sin\left(\frac{\pi}{\alpha}\right)\left(\frac{D\gamma}{\alpha\lambda}\right)^{1/\alpha}\sum_{n,m=0}^{\infty}\frac{(-1)^{n}(x_0/D^{\frac{1}{\alpha}})^n\;\psi_{n,m}(0)}{\gamma\Gamma(n+1)\Gamma(m+1)\Gamma\left(\frac{1}{\alpha}\right)}\;\frac{\alpha\lambda(1-\alpha)}{(n+1)+(m-1)\alpha}\nonumber\\
\mathcal{C}_2^{har}&=&\alpha\sin\left(\frac{\pi}{\alpha}\right)\left(\frac{D\gamma}{\alpha\lambda}\right)^{1/\alpha}\sum_{n,m=0}^{\infty}\frac{(-1)^{n}(x_0/D^{\frac{1}{\alpha}})^n\;\psi_{n,m}(0)}{\Gamma(n+1)\Gamma(m+1)}\left.w'(s)\right|_{s=-(\alpha-1)\lambda/\gamma},\nonumber\\
&&
\end{eqnarray}
with
\begin{equation}
w(s)=\frac{\alpha^2\lambda}{\gamma}\left(\frac{D\gamma}{\alpha\lambda}\right)^{\frac{1}{\alpha}}\frac{s^2\sin\left(\frac{\pi}{\alpha}\right)}{s+(n+m\alpha)\frac{\lambda}{\gamma}} \frac{\Gamma^2(2-\frac{1}{\alpha}+\frac{s\gamma}{\alpha\lambda})}{\Gamma^2\left(1+\frac{s\gamma}{\alpha\lambda}\right)}.
\end{equation}
The absorption time distribution decays exponentially with $(\alpha-1)\lambda/\gamma$ as the rate constant. It correctly predicts the trend observed in Fig. \ref{Fig:first_arrival_har_diff} - larger the force constant, faster the decay. The effect of a finite sink is accounted for in this result as an additive term. This correction for a finite sink is valid only for fairly large sink strengths, i.e. when $1/k_0^2<<1$, while the correction for a finite sink in the previous problems was valid for any sink strength at long times.

\section{Simulation Details}
\label{Sec:simulations}

Extensive Langevin dynamics simulations are performed for L\'{e}vy flights in the presence of a point sink. The underlying equation of motion which is propagated is
\begin{equation}
\label{Eq:langevin_equation}
\dot{x}(t)=-\frac{V'(x)}{m\gamma}+\eta_{\alpha}(t),
\end{equation}
where $V(x)$ is the potential, $m$ is the mass of the particle, $\gamma$ is the friction constant and $\eta_{\alpha}(t)$ is the white, symmetric L\'{e}vy noise acting on the particle. Starting from the above continuous time equation, it is possible to write down its discretized version \cite{chechkin_numerical} as
\begin{equation}
\label{Eq:langevin_discretized}
x_{n+1}-x_{n}=-\frac{V'(x_n)}{m\gamma}\Delta t+\eta_{\alpha,0}(n;\left(D\Delta t\right)^{\frac{1}{\alpha}}),
\end{equation}
where $\Delta t$ is the time-step and $t=n\Delta t$ with $n=0,1,2,...$ $\eta_{\alpha,0}(n; \left(D\Delta t\right)^{\frac{1}{\alpha}})$ is the L\'{e}vy noise acting on the particle at the $n^{th}$ time-step which is drawn from an $\alpha$-stable distribution with zero mean and noise density $\sigma=(D\Delta t)^{1/\alpha}$ whose characteristic function is given by
\begin{equation}
\label{characteristic_function}
\tilde{L}_{\alpha,0}(k)=e^{-D|k|^{\alpha}\Delta t}.
\end{equation}

The aim of the simulations is to model the absorption at the point sink and obtain the absorption time distribution for sinks of various strengths in all the three potentials discussed in the previous section. Mathematically, a point sink is modeled using a delta function as given in Eq. (\ref{Eq:Levy_smoluchowski_sink}). When the particle arrives exactly at $x_s$, it will be absorbed by the sink with a probability $P(k_0)$. However, in an actual simulation, the probability of finding one point, $x_s$, is zero. Therefore,  instead of a delta function sink, the sink is provided with a certain width $w$  and modeled using a window function, and the resulting Fokker-Planck equation is given by
\begin{equation}
 \label{Eq:Levy_smoluchowski_window}
 \frac{\partial P(x,t)}{\partial t}=\left\{-D\left(-\frac{\partial^2}{\partial x^2}\right)^{\alpha/2}+\frac{\partial}{\partial x}\frac{V'(x)}{m\gamma}-\frac{k_0}{w}\left(\Theta\left(x-x_s+\frac{w}{2}\right)-\Theta\left(x-x_s-\frac{w}{2}\right)\right)\right\}P(x,t).
\end{equation}
If the particle arrives within the interval $\left(x_s-\frac{w}{2},x_s+\frac{w}{2}\right)$, the particle is absorbed with the probability $P(k_0)$. (The window function is divided by the width of the window, $w$, to maintain the dimensional correctness.)

The probability of absorption, $P(k_0)$, which is a function of the sink strength can be evaluated by recognizing that the absorption described in Eq. (\ref{Eq:Levy_smoluchowski_window}) is a first-order process with a rate constant $k_0/w$. In a simulation, the particle is propagated in discrete time-steps of size $\Delta t$. Thus, once the particle reaches the interval $\left(x_s-\frac{w}{2},x_s+\frac{w}{2}\right)$, its time of residence is $\Delta t$. The probability of absorption for a first-order process in a time interval $\Delta t$ with a rate-constant $k_0/w$ is
\begin{equation}
\label{Eq:prob_abs}
P(k_0)=1-e^{-\frac{k_0}{w}\Delta t}.
\end{equation}
For a sink of infinite strength, $P(k_0\rightarrow\infty)=1$ and for $k_0=0$, $P(0)=0$ which just implies the absence of a sink.

Using the appropriate forms of the potential in Eq. (\ref{Eq:langevin_discretized}), the absorption time distribution is evaluated for $k_0\rightarrow\infty$ and $k_0=5$. The simulation parameters are the following: $\Delta t=10^{-3}$, $m=1$, $\gamma=1$, and $D=1$. A total of $2\times 10^6$ trajectories are propagated for each simulation. $w=0.01$ and $w=0.003$ are used for $k_0\rightarrow\infty$ and $k_0=5$, respectively. The choice of $w$ is done such that $\sim 70\%$ of the trajectories reach the sink in the time of propagation. The effect of different values of $F$ and $\lambda$ are simulated for the linear and harmonic potentials, respectively. The results of the simulations are plotted in figures giving the absorption time distribution for different potentials and are found to corroborate very closely with the numerically calculated Laplace inverse of Eq. (\ref{Eq:first_arrival_derv}).

\section{Conclusions}

In this paper, we consider the problem of L\'{e}vy flights in the presence of a point sink of \textit{arbitrary strength} in three different potentials: $V(x)=0$, $V(x)=-Fx$ and $V(x)=\lambda x^2/2$, and calculate their corresponding Green's function and the absorption time distribution. We employ the elegant formula given in Eq. (\ref{Eq:Greens_function_laplace}) where the Laplace transform of the Green's function in the presence of a sink is given in terms of the Laplace transform of the Green's function in the absence of a sink.

The asymptotic behaviour of the absorption time distributions which are obtained analytically are the following:
\begin{enumerate}
\item{$V(x)=0$:
\begin{equation}
p_{abs}^{free}(t)\sim\left(\mathcal{C}_1^{free}+\frac{\mathcal{C}_2^{free}}{k_0}\right)\frac{1}{t^{2-\frac{1}{\alpha}}}.\end{equation}
This result is valid in the region $1<\alpha\leq 2$ and arbitrary position of the sink, and shows $\sim t^{-2+\frac{1}{\alpha}}$ power-law behaviour at long times which is in agreement with $\sim t^{-\frac{3}{2}}$ behaviour in the case of Brownian motion ($\alpha=2$).}
\item{$V(x)=-F x$:
\begin{equation}
p_{abs}^{lin}(t)\sim\left(\mathcal{C}_1^{free}+\frac{\mathcal{C}_2^{free}}{k_0}\right)\frac{1}{t^{2-\frac{1}{\alpha}}}+\frac{F}{\gamma}(x_s-x_0)\left(\frac{\mathcal{C}_3^{lin}}{t^{2-\frac{1}{\alpha}}}-\frac{\mathcal{B}^{lin}}{t^{\frac{2}{\alpha}}}\right),
\end{equation}
which is valid for $1<\alpha\leq 2$, $F<<1$ and arbitrary sink position. $\mathcal{B}^{lin}$ vanishes exactly for $\alpha=2$, reproducing the $\sim t^{-\frac{3}{2}}$ decay of the $F$-dependent term for Brownian motion. For $\frac{3}{2}<\alpha\leq 2$, the power-law decay begins as $\sim t^{-2+\frac{1}{\alpha}}$ and  $\sim t^{-\frac{2}{\alpha}}$  becomes dominant at very long times. When $1<\alpha<\frac{3}{2}$, the trend is reversed where the asymptotic behaviour is  $\sim t^{-\frac{2}{\alpha}}$ initially, followed by $\sim t^{-2+\frac{1}{\alpha}}$ for very large $t$. The position of the sink with respect to the initial position is crucial in determining the effect of the potential on the long-time limit of $p_{abs}^{lin}(t)$.}
\item{$V(x)=\lambda x^2/2$:
\begin{equation}
p_{abs}^{har}(t)\sim\left(\mathcal{C}_1^{har}+\frac{\mathcal{C}_2^{har}}{k_0}\right)e^{-(\alpha-1)\frac{\lambda t}{\gamma}},
\end{equation}
which is valid for $1<\alpha\leq 2$, $1/k_0^2<<1$ and sink placed at the origin ($x_s=0$), and confirms with $\sim e^{-\frac{\lambda t}{\gamma}}$ decay for Brownian motion. According to this result, a larger force constant results in a faster decay of $p_{abs}^{har}(t)$ for a given value of $\alpha$.}
\end{enumerate}

Explicit expressions for the coefficients of decay are provided in the corresponding sections of the paper. The absorption time distributions in all the three potentials for $0<\alpha\leq 1$ are zero at all times going to show that such a L\'{e}vy flight will never find a point target \cite{search_levy_brown}. As for the effect of a sink of finite strength, it can be concluded that the value of $k_0$ does not alter the nature of decay of the absorption time distribution, i.e. the exponent in a power-law or an exponential decay depends solely on the value of $\alpha$. With all things being equal, a finite sink strength will lead to a larger multiplicative constant in the asymptotic behaviour.  The above results can be very useful in evolving L\'{e}vy search strategies in these potentials at long-times.

\section{Acknowledgement}
Funding for this work was provided by the Simons Centre for Living Machines, NCBS, Bangalore. I would like to thank Prof. K. L. Sebastian for all the discussions and suggestions, and Prof. Madan Rao for his support.

\appendix

\section{Green's function solution for a reaction-diffusion equation with a finite point sink in the Laplace domain}
\label{App:greens_func}

The reaction-diffusion equation in presence of a finite delta function sink is given by
\begin{equation}
\label{Eq:smoluchowski_laplacian}
\frac{\partial P(x,t)}{\partial t}=\left\{\mathcal{L}-k_0 \delta(x-x_s)\right\}P(x,t),
\end{equation}
where $\mathcal{L}$ could be the Smoluchowski operator for fractional or normal diffusion. In the Laplace domain, the above PDE is given by
\begin{equation}
\label{Eq:smoluchowski_laplacian_laplace}
\left[s-\mathcal{L}+k_0\delta(x-x_s)\right]\tilde{P}(x,s)=P_0(x),
\end{equation}
where $\tilde{P}(x,s)$ is the Laplace transform of $P(x,t)$. The exact solution for this equation can be obtained by writing it in terms of its Green's function \cite{Sebastian_IASc,Sebastian_CPL}, $\mathcal{G}(x,s|x_0)$, as
\begin{equation}
\label{Eq:greens_function_equation}
\left[s-\mathcal{L}+k_0\delta(x-x_s)\right]\mathcal{G}(x,s|x_0)=\delta(x-x_0).
\end{equation}
$\mathcal{G}(x,s|x_0)$ is the propagator in the Laplace domain which gives $\tilde{P}(x,s)$ from the relation
\begin{equation}
\label{Eq:prob_density_greens}
\tilde{P}(x,s)=\int^{\infty}_{-\infty}dx_0\;\mathcal{G}(x,s|x_0)P_0 (x_0).
\end{equation}
The Green's function can be written down in terms bra-ket notation of quantum mechanics as
\begin{equation}
\label{Eq:braket_greens}
\mathcal{G}(x,s|x_0)=\langle x|\left[s-\mathcal{L}+k_0\delta(x-x_s)\right]|x_0\rangle,
\end{equation}
which implicitly assumes the existence of an expansion of the propagator in terms of its eigenfunctions $|x\rangle\left(|x_0\rangle\right)$ and the corresponding eigenvalues. Further, the operator identity
\begin{equation}
\label{Eq:operator_identity}
\left[s-\mathcal{L}+k_0\delta(x-x_s)\right]^{-1}=\left[s-\mathcal{L}\right]^{-1}-\left[s-\mathcal{L}\right]^{-1}\left(k_0\delta(x-x_s)\right)\left[s-\mathcal{L}+k_0\delta(x-x_s)\right]^{-1},
\end{equation}
can be employed in the bra-ket notation for the Green's function to obtain
\begin{eqnarray}
\label{Eq:operator_identity_greens}
\mathcal{G}(x,s|x_0)&=&\langle x|\left[s-\mathcal{L}+k_0\delta(x-x_s)\right]^{-1}|x_0\rangle\nonumber\\
&=&\langle x|\left[s-\mathcal{L}\right]^{-1}|x_0\rangle-\langle x|\left[s-\mathcal{L}\right]^{-1}\left(k_0\delta(x-x_s)\right)\left[s-\mathcal{L}+k_0\delta(x-x_s)\right]^{-1}|x_0\rangle.\nonumber\\
&&
\end{eqnarray}
It is necessary to note that the
\begin{equation}
\label{Eq:unperturbed_greens}
\mathcal{G}_0(x,s|x_0)=\langle x|\left[s-\mathcal{L}\right]^{-1}|x_0\rangle,
\end{equation}
is the solution of the Green's function in the absence of a sink. Upon inserting the resolution of identity, $I=\int^{\infty}_{-\infty}dy |y\rangle \langle y|$, in the last term of Eq. (\ref{Eq:operator_identity_greens}) between the two inverses, we get
\begin{equation}
\label{Eq:operator_identity_greens_1}
\mathcal{G}(x,s|x_0)=\mathcal{G}_0(x,s|x_0)-k_0\int^{\infty}_{-\infty}dy\;\mathcal{G}_0(x,s|y)\delta(y-x_s)\mathcal{G}(y,s|x_0).
\end{equation}
It is trivial to perform the integral over $y$ to obtain
\begin{equation}
\label{Eq:operator_identity_greens_2}
\mathcal{G}(x,s|x_0)=\mathcal{G}_0(x,s|x_0)-k_0 \mathcal{G}_0(x,s|x_s)\mathcal{G}(x_s,s|x_0).
\end{equation}
Using this equation we can solve for $\mathcal{G}(x_s,s|x_0)$ and further insert the solution back in Eq. (\ref{Eq:operator_identity_greens_2}) to obtain the relation given in Eq. (\ref{Eq:Greens_function_laplace}) which is
\begin{equation}
\label{Eq:operator_identity_greens_3}
 \mathcal{G}(x,s|x_0)=\mathcal{G}_0(x,s|x_0)-\frac{k_0 \mathcal{G}_0(x,s|x_s)\mathcal{G}_0(x_s,s|x_0)}{1+k_0\mathcal{G}_0(x_s,s|x_s)}.
\end{equation}

\section{Laplace transform of a Fox H-function}
\label{appendix_1}

The Laplace transform of an H-function can written in terms of another H-function which is given by \cite{Mathai}
\begin{eqnarray}
 \label{Eq:h_function_laplace}
 L\left\{t^{\rho-1}H^{m,n}_{p,q}\left[at^{\sigma}\left|\begin{matrix}
(a_p,A_p) \\
(b_q,B_q)
                                                                                                                    \end{matrix}
\right.\right];s\right\}&=&\int^{\infty}_0 dt \;e^{-s t}\;t^{\rho-1}H^{m,n}_{p,q}\left[at^{\sigma}\left|\begin{matrix}
(a_p,A_p) \\
(b_q,B_q)
                                                                                                                    \end{matrix}
\right.\right]\nonumber\\
&=&s^{-\rho}H^{m,n+1}_{p+1,q}\left[as^{-\sigma}\left|\begin{matrix}
(1-\rho,\sigma),(a_p,A_p) \\
(b_q,B_q)
                                                                                                                    \end{matrix}
\right.\right].
\end{eqnarray}
The Laplace inverse of an H-function is
\begin{eqnarray}
 \label{Eq:h_function_laplace_inverse}
 L^{-1}\left\{s^{-\rho}H^{m,n}_{p,q}\left[as^{\sigma}\left|\begin{matrix}
(a_p,A_p) \\
(b_q,B_q)
                                                                                                                    \end{matrix}
\right.\right];t\right\}&=&t^{\rho-1}H^{m,n}_{p+1,q}\left[at^{-\sigma}\left|\begin{matrix}
(a_p,A_p),(\rho,\sigma), \\
(b_q,B_q)
                                                                                                                    \end{matrix}
\right.\right].
\end{eqnarray}
The above equations are valid for $\rho, a, s \in\mathcal{C}$ and $\sigma>0$.

\section{Long-time behaviour of $p_{abs}^{free}(t)$}
\label{appendix_2}

Starting from the $\tilde{p}^{free}_{abs}(s)$ given in Eq. (\ref{Eq:first_arrival_free_levy}), we are required to obtain its small-$s$ behaviour. Using the large argument expansion (valid for small-s values) of the H-function \cite{Mathai}, we obtain
\begin{eqnarray}
\label{Eq:appendix_free_laplace}
\tilde{p}^{free}_{abs}(s)&=&\frac{D^{1/\alpha}\;s^{-1/\alpha}\;\sin(\pi/\alpha)}{|x_s-x_0|\left(1+\frac{\alpha D^{\frac{1}{\alpha}}s^{1-\frac{1}{\alpha}}}{k_0\;\csc(\pi/\alpha)}\right)}\;H^{1,2}_{3,2}\left[\frac{D^{1/\alpha}s^{-1/\alpha}}{|x_s-x_0|}\left|\begin{matrix}
(0,1/\alpha),(0,1),(0,1/2) \\
(0,1/\alpha),(0,1/2)
                                                                                                                    \end{matrix}
\right.\right]\nonumber\\
&=&\left(1+\frac{\alpha\Gamma(1-\alpha)\sin(\pi/\alpha)}{\Gamma\left(1-\frac{\alpha}{2}\right)\Gamma\left(\frac{\alpha}{2}\right)}\frac{\left|x_s-x_0\right|^{\alpha-1}}{D^{1-\frac{1}{\alpha}}}s^{1-\frac{1}{\alpha}}+....\right)\left(1-\frac{\alpha D^{\frac{1}{\alpha}}s^{1-\frac{1}{\alpha}}}{k_0\;\csc(\pi/\alpha)}+...\right)\nonumber\\
&=&1+\left(\frac{\alpha\Gamma\left(1-\alpha\right)\sin\left(\frac{\pi}{\alpha}\right)}{\Gamma\left(1-\frac{\alpha}{2}\right)\Gamma\left(\frac{\alpha}{2}\right)}\;\frac{|x_s-x_0|^{\alpha-1}}{D^{1-\frac{1}{\alpha}}}-\frac{\alpha D^{\frac{1}{\alpha}}s^{1-\frac{1}{\alpha}}}{k_0\;\csc(\pi/\alpha)}\right)s^{1-\frac{1}{\alpha}}+...
\end{eqnarray}
In order to obtain the Laplace inverse, one can complete the above expansion to an exponential and identify the exponential with an H-function, i.e. $e^{-x}=H^{1,0}_{0,1}\left[x\left|\begin{matrix}-\\(0,1)\end{matrix}\right.\right]$, which can be inverted using Eq. (\ref{Eq:h_function_laplace_inverse}) \cite{metzler_search}.  The long-time behaviour of $p_{abs}^{free}(t)$ is evaluated to be
\begin{eqnarray}
\label{Eq:appendix_free_time}
p_{abs}^{free}(t)&\sim&\left(\frac{\alpha\Gamma\left(1-\alpha\right)\sin\left(\frac{\pi}{\alpha}\right)}{\Gamma\left(1-\frac{\alpha}{2}\right)\Gamma\left(\frac{\alpha}{2}\right)}\;\frac{|x_s-x_0|^{\alpha-1}}{D^{1-\frac{1}{\alpha}}}-\frac{\alpha D^{\frac{1}{\alpha}}s^{1-
\frac{1}{\alpha}}}{k_0\;\csc(\pi/\alpha)}\right)\frac{1}{\Gamma\left(\frac{1}{\alpha}-1\right)}\frac{1}{t^{2-\frac{1}{\alpha}}}\nonumber\\
&&\nonumber\\
&&\mbox{upon simple algebraic manipulations using the identity\;}\Gamma(1+x)=x\Gamma(x)\nonumber\\
&&\nonumber\\
&\sim&\left(\frac{\alpha\Gamma\left(2-\alpha\right)\Gamma\left(2-\frac{1}{\alpha}\right)\sin\left(\frac{\pi\alpha}{2}\right)\sin^2\left(\frac{\pi}{\alpha}\right)}{\pi^2\left(\alpha-1\right)}\;\frac{|x_s-x_0|^{\alpha-1}}{D^{1-\frac{1}{\alpha}}}+\frac{(\alpha-1) D^{\frac{1}{\alpha}}\sin\left(\frac{\pi}{\alpha}\right)}{k_0\Gamma\left(\frac{1}{\alpha}\right)}\right)\frac{1}{t^{2-\frac{1}{\alpha}}}.\nonumber\\
&&
\end{eqnarray}

\section{Long-time behaviour of $p_{abs}^{lin}(t)$ for $F<<1$}
\label{appendix_3}

The Laplace transform of the propagator is given by
\begin{eqnarray}
\label{Eq:laplace_lin_convergence}
\mathcal{G}_0^{lin}(x,s|x_0)&=&\frac{1}{2\pi}\int^{\infty}_0 dt e^{-st} \int^{\infty}_{-\infty}dp e^{-D t|p|^{\alpha}} e^{i p \left(x-x_0-F t/\gamma\right)}\nonumber\\
&&\mbox{performing the integral over }t\nonumber\\
&=&\frac{1}{\pi}\int^{\infty}_{0}dp\;\frac{\cos(p(x-x_0))\left(Dp^{\alpha}+s\right)}{\left(Dp^{\alpha}+s\right)^2+(F/\gamma)^2p^2}+\frac{F/\gamma}{\pi}\int^{\infty}_{0}dp\frac{p\sin(p(x-x_0))}{\left(Dp^{\alpha}+s\right)^2+(F/\gamma)^2p^2}.\nonumber\\
&&
\end{eqnarray}
In the limit $F<<1$, $F^2$ will be negligibly small that it can be neglected. Correspondingly, the Laplace transform of the Green's functions will be
\begin{eqnarray}
\label{Eq:small_force_laplace}
\mathcal{G}_0^{lin}(x,s|x_0)&\approx& \frac{1}{\pi}\int^{\infty}_{0}dp\frac{\cos(p(x-x_0))}{\left(Dp^{\alpha}+s\right)}+\frac{F/\gamma}{\pi}\int^{\infty}_{0}dp\frac{p\sin(p(x-x_0))}{\left(Dp^{\alpha}+s\right)^2}\\
&&\mbox{and}\nonumber\\
\label{Eq:small_force_laplace_loop}
\mathcal{G}_0^{lin}(x,s|x)&\approx& \frac{1}{\pi}\int^{\infty}_{0}dp\frac{1}{\left(Dp^{\alpha}+s\right)}=\frac{\csc(\pi/\alpha)}{\alpha D^{\frac{1}{\alpha}}s^{1-\frac{1}{\alpha}}}.
\end{eqnarray}
The Laplace transform of the absorption time distribution when $F<<1$ is therefore,
\begin{eqnarray}
\label{Eq:first_arrival_lin_derv}
\tilde{p}_{abs}^{lin}(s)&=&\frac{\alpha D^{\frac{1}{\alpha}}s^{1-\frac{1}{\alpha}}}{\pi\csc(\pi/\alpha)\left(1+\frac{\alpha D^{\frac{1}{\alpha}}s^{1-\frac{1}{\alpha}}}{k_0\csc(\pi/\alpha)}\right)}\left(\int^{\infty}_{0}dp\frac{\cos(p(x_s-x_0))}{\left(Dp^{\alpha}+s\right)}+\frac{F}{\gamma}\int^{\infty}_{0}dp\frac{p\sin(p(x_s-x_0))}{\left(Dp^{\alpha}+s\right)^2}\right)\nonumber\\
&=&\frac{\alpha D^{\frac{1}{\alpha}}s^{1-\frac{1}{\alpha}}}{\pi\csc(\pi/\alpha)}\left(\int^{\infty}_{0}dp\frac{\cos(p(x_s-x_0))}{\left(Dp^{\alpha}+s\right)}\right)\left(1-\frac{\alpha D^{\frac{1}{\alpha}}s^{1-\frac{1}{\alpha}}}{k_0\csc(\pi/\alpha)}+..\right)\nonumber\\
&&+\frac{F}{\gamma}\frac{\alpha D^{\frac{1}{\alpha}}s^{1-\frac{1}{\alpha}}}{\pi\csc(\pi/\alpha)}\left(\int^{\infty}_{0}dp\frac{p\sin(p(x_s-x_0))}{\left(Dp^{\alpha}+s\right)^2}\right)\left(1-\frac{\alpha D^{\frac{1}{\alpha}}s^{1-\frac{1}{\alpha}}}{k_0\csc(\pi/\alpha)}+..\right).
\end{eqnarray}
In the above expression, the terms independent of $F$ are identical to those in the free L\'{e}vy flight problem and we can use the analysis done in Appendix \ref{appendix_2} to obtain their small-$s$ limit.

The integral in the force-dependent term in Eq. (\ref{Eq:first_arrival_lin_derv}), $\int^{\infty}_{0}dp\frac{p\sin(p(x_s-x_0))}{\left(Dp^{\alpha}+s\right)^2}$, can be evaluated by writing the $\sin$ function as a contour integral:
\begin{eqnarray}
\label{Eq:small_force_sin}
\frac{1}{\pi}\int^{\infty}_{0}dp\frac{p\sin(p(x_s-x_0))}{\left(Dp^{\alpha}+s\right)^2}&=&\frac{(x_s-x_0)}{\pi|x_s-x_0|}\int^{\infty}_{0}dp\frac{p}{\left(Dp^{\alpha}+s\right)^2}\frac{1}{2i}\int^{c+i\infty}_{c-i\infty}dk\;\frac{\Gamma(k)}{\Gamma\left(\frac{k}{2}\right)\Gamma\left(1-\frac{k}{2}\right)}\left(p|x_s-x_0|\right)^{-k}\nonumber\\
&&\nonumber\\
&=&\frac{(x_s-x_0)}{2\pi i\alpha^2}\int^{c+i\infty}_{c-i\infty}dk\;U(k)\;s^{\frac{2-2\alpha-k}{\alpha}},\nonumber
\end{eqnarray}
where
\begin{equation}
U(k)=\frac{\Gamma(k)\Gamma(1-\frac{2-k}{\alpha})\Gamma(\frac{2-k}{\alpha})}{\Gamma\left(\frac{k}{2}\right)\Gamma\left(1-\frac{k}{2}\right)}\frac{(k+\alpha-2)|x_s-x_0|^{-k-1}}{D^{\frac{2-k}{\alpha}}}.
\end{equation}
There are two sets of poles for the above contour integral upon closing the contour on the left-hand side using a semicircle of radius $R\rightarrow\infty$ which are
\begin{eqnarray}
k_1^{*}(n)&=&-n\;\;\forall\;n\in\mathbb{N},\;\;\;\mbox{arising from }\Gamma(k)\\
k_2^{*}(n)&=&2-\alpha-n\alpha\;\;\forall\;n\in\mathbb{N},\;\;\;\mbox{arising from }\Gamma\left(1-\frac{2-k}{\alpha}\right).
\end{eqnarray}
The contour integral can be evaluated as
\begin{equation}
\frac{1}{2\pi i}\int^{c+i\infty}_{c-i\infty}dk\;U(k)\;s^{\frac{2-2\alpha-k}{\alpha}}=\sum^{\infty}_{n=0}\mbox{Res}\left[U\left(k_1^{*}(n)\right)s^{\frac{2-2\alpha-k_1^{*}(n)}{\alpha}}\right]+\sum^{\infty}_{n=0}\mbox{Res}\left[U\left(k_2^{*}(n)\right)s^{\frac{2-2\alpha-k_2^{*}(n)}{\alpha}}\right].\nonumber\\
\end{equation}
Using the above analysis, one can find that lowest order terms in the small-$s$ expansion of $\tilde{p}_{abs}^{lin}(s)$ to be
\begin{eqnarray}
\tilde{p}_{abs}^{lin}(s)&=&1+\frac{F}{\gamma}(x-x_0)\frac{\Gamma\left(2-\frac{3}{\alpha}\right)\Gamma\left(\frac{3}{\alpha}\right)\sin\left(\frac{\pi}{\alpha}\right)}{\pi D^{\frac{2}{\alpha}}}s^{-1+\frac{2}{\alpha}}+\left(\frac{\alpha\Gamma\left(1-\alpha\right)\sin\left(\frac{\pi}{\alpha}\right)}{\Gamma\left(1-\frac{\alpha}{2}\right)\Gamma\left(\frac{\alpha}{2}\right)}\;\frac{|x_s-x_0|^{\alpha-1}}{D^{1-\frac{1}{\alpha}}}\right.\nonumber\\
&&\left.-\frac{\alpha D^{\frac{1}{\alpha}}}{k_0\;\csc(\pi/\alpha)}+\frac{F}{\gamma}(x-x_0)\frac{\alpha \Gamma(2-2\alpha)\sin\left(\frac{\pi}{\alpha}\right)\sin(\pi\alpha)}{\pi}\frac{|x-x_0|^{2\alpha-3}}{D^{2-\frac{1}{\alpha}}}\right)s^{1-\frac{1}{\alpha}}+...\nonumber\\
&&
\end{eqnarray}
The Laplace inverse of the above equation giving the long-time behaviour of the absorption time distribution after simple algebraic manipulations using the identity $\Gamma(1+x)=x\Gamma(x)$ is
\begin{eqnarray}
p_{abs}^{lin}(t)&\sim&\left\{\frac{(\alpha-1) |x_s-x_0|^{\alpha-1}}{D^{1-\frac{1}{\alpha}}}\left(\frac{\pi\;\Gamma(3-\alpha)}{2\alpha\;\Gamma\left(2-\frac{1}{\alpha}\right)\Gamma\left(2-\frac{\alpha}{2}\right)\Gamma^2\left(\frac{1}{\alpha}\right)\Gamma\left(\frac{\alpha}{2}\right)}\right)+\frac{1}{k_0}\frac{(\alpha-1)^2\;D^{\frac{1}{\alpha}}\pi}{\alpha\Gamma\left(2-\frac{1}{\alpha}\right)\Gamma^2\left(\frac{1}{\alpha}\right)}\right.\nonumber\\
&&+\frac{F}{\gamma}(x_s-x_0)\left.\frac{(\alpha-1)^2|x_s-x_0|^{2\alpha-3}}{(2\alpha-3)D^{2-\frac{1}{\alpha}}}\left(\frac{\pi\Gamma(5-2\alpha)}{4\alpha\;\Gamma(\alpha)\Gamma(3-\alpha)\Gamma\left(2-\frac{1}{\alpha}\right)\Gamma^2\left(\frac{1}{\alpha}\right)}\right)\right\}\frac{1}{t^{2-\frac{1}{\alpha}}}\nonumber\\
&&-\left\{\frac{F}{\gamma}(x_s-x_0)\frac{(2-\alpha)(\alpha-1)}{(2\alpha-3)D^{\frac{2}{\alpha}}}\left(\frac{2\;\Gamma\left(4-\frac{3}{\alpha}\right)\Gamma\left(\frac{3}{\alpha}\right)}{3\alpha\;\Gamma\left(3-\frac{2}{\alpha}\right)\Gamma\left(2-\frac{1}{\alpha}\right)\Gamma\left(\frac{1}{\alpha}\right)}\right)\right\}\frac{1}{t^{\frac{2}{\alpha}}}.
\end{eqnarray}

\section{Harmonic potential}
\label{appendix_4}

\textbf{Laplace transform of $G_0^{har}(x,t|x)$}

The Laplace transform, $\mathcal{G}_0^{har}(x,s|x_0)$, given in Eq. (\ref{Eq:Levy_harmonic_laplace}) is valid for $0<\alpha\leq 2$. The Laplace transform of the loop propagator with the initial and final position placed at the origin is
\begin{eqnarray}
\mathcal{G}_0^{har}(0,s|0)&=&\frac{1}{\pi}\int^{\infty}_0 dt\;e^{-st}\int^{\infty}_0dp\;e^{-\frac{D\gamma}{\alpha\lambda}p^{\alpha}\left(1-e^{\alpha\lambda t/\gamma}\right)}\nonumber\\
&=&\left(\frac{\alpha\lambda}{D\gamma}\right)^{1/\alpha}\frac{\Gamma\left(1+\frac{1}{\alpha}\right)}{\pi}\int^{\infty}_0dt\;e^{-s t}\left(1-e^{-\frac{\alpha\lambda t}{\gamma}}\right)^{-1/\alpha}\nonumber\\
&&\mbox{making a change of variable such that }g=\frac{1}{1-e^{-\alpha\lambda t/\gamma}}-1\nonumber\\
\label{Eq:har_int_conv}
&=&\left(\frac{\alpha\lambda}{D\gamma}\right)^{1/\alpha}\frac{\gamma\Gamma\left(1+\frac{1}{\alpha}\right)}{\pi\alpha\lambda}\int^{\infty}_{0}dg\;\frac{g^{\frac{s\gamma}{\alpha\lambda}-1}}{(1+g)^{\left(2-\frac{1}{\alpha}\right)+\left(\frac{s\gamma}{\alpha\lambda}-1\right)}}\\
&=&\left(\frac{\alpha\lambda}{D\gamma}\right)^{1/\alpha}\frac{\csc\left(\frac{\pi}{\alpha}\right)}{s\alpha}\frac{\Gamma\left(1+\frac{s\gamma}{\alpha\lambda}\right)}{\Gamma\left(1-\frac{1}{\alpha}+\frac{s\gamma}{\alpha\lambda}\right)}.
\end{eqnarray}
It can be easily seen that Eq. (\ref{Eq:har_int_conv}) converges only $1<\alpha\leq 2$.

\textbf{Long-time behaviour of $p_{abs}^{har}(t)$}

The absorption time distribution in a harmonic potential with the sink placed at the origin can be written in terms of a contour integral as given in Eq. (\ref{Eq:Levy_harmonic_contour}) as
\begin{eqnarray}
p_{abs}^{har}(t)&=&\frac{1}{2\pi i}\int^{c+i\infty}_{c-i\infty} ds\;e^{st}\frac{k_0 \mathcal{G}_0^{har}(0,s|x_0)}{1+k_0 \mathcal{G}_0^{har}(0,s|0)}\nonumber\\
&=&\alpha\sin\left(\frac{\pi}{\alpha}\right)\left(\frac{D\gamma}{\alpha\lambda}\right)^{1/\alpha}\sum_{n,m=0}^{\infty}\frac{(-x_0/D^{\frac{1}{\alpha}})^n}{\Gamma(n+1)\Gamma(m+1)}\psi_{n,m}(0) \frac{1}{2\pi i}\int^{c+i\infty}_{c-i\infty}ds\;e^{s t}\;W^{n,m}(s),\nonumber\\
&&
\end{eqnarray}
where
\begin{equation}
W^{n,m}(s)=\frac{s}{s+(n+m\alpha)\frac{\lambda}{\gamma}} \frac{\Gamma(1-\frac{1}{\alpha}+\frac{s\gamma}{\alpha\lambda})}{\Gamma\left(1+\frac{s\gamma}{\alpha\lambda}\right)}\frac{1}{1+\frac{s\alpha\sin(\pi/\alpha)}{k_0}\left(\frac{D\gamma}{\alpha\lambda}\right)^{1/\alpha}\frac{\Gamma\left(1-\frac{1}{\alpha}+\frac{s\gamma}{\alpha\lambda}\right)}{\Gamma\left(1+\frac{s\gamma}{\alpha\lambda}\right)}}
\end{equation}
When the sink strength is fairly large, $W^{n,m}(s)$ can be written as
\begin{equation}
\label{Eq:integrand_expansion_har}
W^{n,m}(s)\approx W_1^{n,m}(s)+\frac{W_2^{n,m}(s)}{k_0},
\end{equation}
where
\begin{eqnarray}
W_1^{n,m}(s)&=&\frac{s}{s+(n+m\alpha)\frac{\lambda}{\gamma}} \frac{\Gamma(1-\frac{1}{\alpha}+\frac{s\gamma}{\alpha\lambda})}{\Gamma\left(1+\frac{s\gamma}{\alpha\lambda}\right)},\nonumber\\
W_2^{n,m}(s)&=&-\left(\frac{D\gamma}{\alpha\lambda}\right)^{\frac{1}{\alpha}}\frac{s^2\alpha\sin\left(\frac{\pi}{\alpha}\right)}{s+(n+m\alpha)\frac{\lambda}{\gamma}} \frac{\Gamma^2(1-\frac{1}{\alpha}+\frac{s\gamma}{\alpha\lambda})}{\Gamma^2\left(1+\frac{s\gamma}{\alpha\lambda}\right)}.
\end{eqnarray}
In the approximation in Eq. (\ref{Eq:integrand_expansion_har}), all the terms with $1/k_0^n$ where $n\geq 2$ are neglected.
The poles of $W_1^{n,m}(s)$ and $W_2^{n,m}(s)$ are situated at
\begin{eqnarray}
\label{poles}
s_1^{*}(n,m)&=&-(n+m\alpha)\frac{\lambda}{\gamma}\;\;\forall n,m\in \mathbb{N},\;\;\mbox{arising from }\left(s+(n+m\alpha)\frac{\lambda}{\gamma}\right)^{-1}\nonumber\\
s_2^{*}(k)&=&-\frac{\lambda}{\gamma}(\alpha-1+k\alpha)\;\;\forall k\in\mathbb{N} \;\;\mbox{arising from }\Gamma\left(1-\frac{1}{\alpha}+\frac{s\gamma}{\alpha\lambda}\right).
\end{eqnarray}
It may be noted that the poles at $s_2^{*}(k)$ are simple poles for $W_1^{n,m}(s)$, while these are poles of order $2$ for $W_2^{n,m}(s)$.
Calculation of residues at the poles results in
\begin{eqnarray}
\frac{1}{2\pi i}\int^{c+i\infty}_{c-i\infty}ds\;e^{s t}\;W^{n,m}(s)&\approx&\mbox{Res}\left[e^{s_1^{*}(n,m)}W_1^{n,m}\left(s_1^{*}(n,m)\right)\right]+\mbox{Res}\left[e^{s_1^{*}(n,m)}W_2^{n,m}\left(s_1^{*}(n,m)\right)\right]\nonumber\\
&&+\sum^{\infty}_{k=0} \mbox{Res}\left[e^{s_2^{*}(k)}W_1^{n,m}\left(s_2^{*}(k)\right)\right]+\sum^{\infty}_{k=0} \mbox{Res}\left[e^{s_2^{*}(k)}W_2^{n,m}\left(s_2^{*}(k)\right)\right].\nonumber\\
&&
\end{eqnarray}
One can infer that evaluation of the residue at $s_2^{*}(0)=-(\alpha-1)\lambda/\gamma$ leads to the smallest power in the exponential decay. This is the term which will dominate at long times and hence, it is sufficient to evaluate the residue at this particular pole for this analysis.
\begin{eqnarray}
\mbox{Res}\left[e^{s_2^{*}(0)}W_1^{n,m}\left(s_2^{*}(0)\right)\right]&=&\frac{e^{-(\alpha-1)\lambda t/\gamma}}{\gamma\Gamma\left(\frac{1}{\alpha}\right)}\frac{\alpha\lambda(1-\alpha)}{(n+1)+(m-1)\alpha},\nonumber\\
\mbox{Res}\left[e^{s_2^{*}(0)}W_2^{n,m}\left(s_2^{*}(0)\right)\right]&=& e^{-(\alpha-1)\lambda t/\gamma}\;\left.w'(s)\right|_{s=-\frac{(\alpha-1)\lambda}{\gamma}},
\end{eqnarray}
where
\begin{equation}
w(s)=\frac{\alpha^2\lambda}{\gamma}\left(\frac{D\gamma}{\alpha\lambda}\right)^{\frac{1}{\alpha}}\frac{s^2\sin\left(\frac{\pi}{\alpha}\right)}{s+(n+m\alpha)\frac{\lambda}{\gamma}} \frac{\Gamma^2(2-\frac{1}{\alpha}+\frac{s\gamma}{\alpha\lambda})}{\Gamma^2\left(1+\frac{s\gamma}{\alpha\lambda}\right)}.
\end{equation}
Therefore, asymptotic behaviour of $p_{abs}^{har}(t)$ when $1/k_0^2<<1$ is
\begin{eqnarray}
p_{abs}^{har}(t)&\sim& e^{-(\alpha-1)\lambda t/\gamma}\left\{\alpha\sin\left(\frac{\pi}{\alpha}\right)\left(\frac{D\gamma}{\alpha\lambda}\right)^{1/\alpha}\sum_{n,m=0}^{\infty}\frac{(-x_0/D^{\frac{1}{\alpha}})^n}{\Gamma(n+1)\Gamma(m+1)}\psi_{n,m}(0)\right.\nonumber\\
&&\times\left. \left(\frac{1}{\gamma\Gamma\left(\frac{1}{\alpha}\right)}\frac{\alpha\lambda(1-\alpha)}{(n+1)+(m-1)\alpha}+\frac{w'(s_2^{*}(0))}{k_0}\right)\right\}.
\end{eqnarray}


\begin{thebibliography}{32}
\expandafter\ifx\csname natexlab\endcsname\relax\def\natexlab#1{#1}\fi
\expandafter\ifx\csname bibnamefont\endcsname\relax
  \def\bibnamefont#1{#1}\fi
\expandafter\ifx\csname bibfnamefont\endcsname\relax
  \def\bibfnamefont#1{#1}\fi
\expandafter\ifx\csname citenamefont\endcsname\relax
  \def\citenamefont#1{#1}\fi
\expandafter\ifx\csname url\endcsname\relax
  \def\url#1{\texttt{#1}}\fi
\expandafter\ifx\csname urlprefix\endcsname\relax\def\urlprefix{URL }\fi
\providecommand{\bibinfo}[2]{#2}
\providecommand{\eprint}[2][]{\url{#2}}

\bibitem[{\citenamefont{Redner}(2001)}]{Redner}
\bibinfo{author}{\bibfnamefont{S.}~\bibnamefont{Redner}},
  \emph{\bibinfo{title}{A Guide to First-Passage Processes}}
  (\bibinfo{publisher}{Cambridge University Press}, \bibinfo{year}{2001}).

\bibitem[{\citenamefont{Sebastian}(1994)}]{Sebastian_IASc}
\bibinfo{author}{\bibfnamefont{K.~L.} \bibnamefont{Sebastian}},
  \bibinfo{journal}{Proc. Indian Acad. Sci. (Chem. Sci.)}
  \textbf{\bibinfo{volume}{106}}, \bibinfo{pages}{493} (\bibinfo{year}{1994}).

\bibitem[{\citenamefont{Sumi and Marcus}(1986)}]{Sumi_marcus}
\bibinfo{author}{\bibfnamefont{H.}~\bibnamefont{Sumi}} \bibnamefont{and}
  \bibinfo{author}{\bibfnamefont{R.~A.} \bibnamefont{Marcus}},
  \bibinfo{journal}{J. Chem. Phys.} \textbf{\bibinfo{volume}{84}},
  \bibinfo{pages}{4894} (\bibinfo{year}{1986}).

\bibitem[{\citenamefont{Bagchi and Fleming}(1990)}]{Bagchi}
\bibinfo{author}{\bibfnamefont{B.}~\bibnamefont{Bagchi}} \bibnamefont{and}
  \bibinfo{author}{\bibfnamefont{G.~R.} \bibnamefont{Fleming}},
  \bibinfo{journal}{J. Phys. Chem.} \textbf{\bibinfo{volume}{94}},
  \bibinfo{pages}{9} (\bibinfo{year}{1990}).

\bibitem[{\citenamefont{Chakravarthi and Sebastian}(1993)}]{Sebastian_CPL}
\bibinfo{author}{\bibfnamefont{N.}~\bibnamefont{Chakravarthi}}
  \bibnamefont{and} \bibinfo{author}{\bibfnamefont{K.~L.}
  \bibnamefont{Sebastian}}, \bibinfo{journal}{Chem. Phys. Lett.}
  \textbf{\bibinfo{volume}{206}}, \bibinfo{pages}{496} (\bibinfo{year}{1993}).

\bibitem[{\citenamefont{Risken}(1996)}]{Risken}
\bibinfo{author}{\bibfnamefont{H.}~\bibnamefont{Risken}},
  \emph{\bibinfo{title}{The Fokker-Planck Equation: Methods of Solution and
  Applications}} (\bibinfo{publisher}{Springer-Verlag Berlin Heidelberg New
  York}, \bibinfo{year}{1996}).

\bibitem[{\citenamefont{Metzler and Klafter}(2000)}]{physrepmet}
\bibinfo{author}{\bibfnamefont{R.}~\bibnamefont{Metzler}} \bibnamefont{and}
  \bibinfo{author}{\bibfnamefont{J.}~\bibnamefont{Klafter}},
  \bibinfo{journal}{Physics Reports} \textbf{\bibinfo{volume}{339}},
  \bibinfo{pages}{1} (\bibinfo{year}{2000}).

\bibitem[{\citenamefont{Janakiraman and Sebastian}(2012)}]{deepika_Levy_prop}
\bibinfo{author}{\bibfnamefont{D.}~\bibnamefont{Janakiraman}} \bibnamefont{and}
  \bibinfo{author}{\bibfnamefont{K.~L.} \bibnamefont{Sebastian}},
  \bibinfo{journal}{Phys. Rev. E} \textbf{\bibinfo{volume}{86}},
  \bibinfo{pages}{061105} (\bibinfo{year}{2012}).

\bibitem[{\citenamefont{Chechkin et~al.}(2008)\citenamefont{Chechkin, Metzler,
  Klafter, and Gonchar}}]{chechkinintroduction2008}
\bibinfo{author}{\bibfnamefont{A.~V.} \bibnamefont{Chechkin}},
  \bibinfo{author}{\bibfnamefont{R.}~\bibnamefont{Metzler}},
  \bibinfo{author}{\bibfnamefont{J.}~\bibnamefont{Klafter}}, \bibnamefont{and}
  \bibinfo{author}{\bibfnamefont{V.~Y.} \bibnamefont{Gonchar}},
  \bibinfo{journal}{Anomalous Transport} pp. \bibinfo{pages}{129--162}
  (\bibinfo{year}{2008}).

\bibitem[{\citenamefont{Metzler et~al.}(2000)\citenamefont{Barkai, Metzler, and
  Klafter}}]{MBK_PRE}
\bibinfo{author}{\bibfnamefont{E.}~\bibnamefont{Barkai}},
  \bibinfo{author}{\bibfnamefont{R.}~\bibnamefont{Metzler}}, \bibnamefont{and}
  \bibinfo{author}{\bibfnamefont{J.}~\bibnamefont{Klafter}},
  \bibinfo{journal}{Phys. Rev. E} \textbf{\bibinfo{volume}{61}},
  \bibinfo{pages}{132} (\bibinfo{year}{2000}).

\bibitem[{\citenamefont{Koren et~al.}(2007)\citenamefont{Koren, Lomholt,
  Chechkin, Klafter, and Metzler}}]{leapover_paper}
\bibinfo{author}{\bibfnamefont{T.}~\bibnamefont{Koren}},
  \bibinfo{author}{\bibfnamefont{M.~A.} \bibnamefont{Lomholt}},
  \bibinfo{author}{\bibfnamefont{A.~V.} \bibnamefont{Chechkin}},
  \bibinfo{author}{\bibfnamefont{J.}~\bibnamefont{Klafter}}, \bibnamefont{and}
  \bibinfo{author}{\bibfnamefont{R.}~\bibnamefont{Metzler}},
  \bibinfo{journal}{Phys. Rev. Lett.} \textbf{\bibinfo{volume}{99}},
  \bibinfo{pages}{160602} (\bibinfo{year}{2007}).

\bibitem[{\citenamefont{Garc\'{i}a-Garc\'{i}a
  et~al.}(2012)\citenamefont{Garc\'{i}a-Garc\'{i}a, Rosso, and
  Schehr}}]{levy_half_line}
\bibinfo{author}{\bibfnamefont{R.}~\bibnamefont{Garc\'{i}a-Garc\'{i}a}},
  \bibinfo{author}{\bibfnamefont{A.}~\bibnamefont{Rosso}}, \bibnamefont{and}
  \bibinfo{author}{\bibfnamefont{G.}~\bibnamefont{Schehr}},
  \bibinfo{journal}{Phys. Rev. E} \textbf{\bibinfo{volume}{86}},
  \bibinfo{pages}{011101} (\bibinfo{year}{2012}).

\bibitem[{\citenamefont{Andersen}(1953{\natexlab{a}})}]{Sparre_Andersen_paper1}
\bibinfo{author}{\bibfnamefont{E.~S.} \bibnamefont{Andersen}},
  \bibinfo{journal}{Math. Scand.} \textbf{\bibinfo{volume}{1}},
  \bibinfo{pages}{263} (\bibinfo{year}{1953}{\natexlab{a}}).

\bibitem[{\citenamefont{Andersen}(1953{\natexlab{b}})}]{Sparre_Andersen_paper2}
\bibinfo{author}{\bibfnamefont{E.~S.} \bibnamefont{Andersen}},
  \bibinfo{journal}{Math. Scand.} \textbf{\bibinfo{volume}{2}},
  \bibinfo{pages}{195} (\bibinfo{year}{1953}{\natexlab{b}}).

\bibitem[{\citenamefont{Chechkin et~al.}(2003)\citenamefont{Chechkin, Metzler,
  Gonchar, Klafter, and Tanatarov}}]{metzler_search}
\bibinfo{author}{\bibfnamefont{A.~V.} \bibnamefont{Chechkin}},
  \bibinfo{author}{\bibfnamefont{R.}~\bibnamefont{Metzler}},
  \bibinfo{author}{\bibfnamefont{V.~Y.} \bibnamefont{Gonchar}},
  \bibinfo{author}{\bibfnamefont{J.}~\bibnamefont{Klafter}}, \bibnamefont{and}
  \bibinfo{author}{\bibfnamefont{L.~V.} \bibnamefont{Tanatarov}},
  \bibinfo{journal}{J. Phys. A: Math. Gen.} \textbf{\bibinfo{volume}{36}},
  \bibinfo{pages}{L537} (\bibinfo{year}{2003}).

\bibitem[{\citenamefont{Skaug et~al.}(2013)\citenamefont{Skaug, Mabry, and
  Schwartz}}]{Schwartz1}
\bibinfo{author}{\bibfnamefont{M.~J.} \bibnamefont{Skaug}},
  \bibinfo{author}{\bibfnamefont{J.}~\bibnamefont{Mabry}}, \bibnamefont{and}
  \bibinfo{author}{\bibfnamefont{D.~K.} \bibnamefont{Schwartz}},
  \bibinfo{journal}{Phys. Rev. Lett.} \textbf{\bibinfo{volume}{110}},
  \bibinfo{pages}{256101} (\bibinfo{year}{2013}).

\bibitem[{\citenamefont{Skaug et~al.}(2014)\citenamefont{Skaug, Mabry, and
  Schwartz}}]{Schwartz2}
\bibinfo{author}{\bibfnamefont{M.~J.} \bibnamefont{Skaug}},
  \bibinfo{author}{\bibfnamefont{J.}~\bibnamefont{Mabry}}, \bibnamefont{and}
  \bibinfo{author}{\bibfnamefont{D.~K.} \bibnamefont{Schwartz}},
  \bibinfo{journal}{J. Am. Chem. Soc.} \textbf{\bibinfo{volume}{136}},
  \bibinfo{pages}{1327} (\bibinfo{year}{2014}).

\bibitem[{\citenamefont{Yu et~al.}(2013)\citenamefont{Yu, Guan, Chen, C.Bae,
  and Granick}}]{Granick}
\bibinfo{author}{\bibfnamefont{C.}~\bibnamefont{Yu}},
  \bibinfo{author}{\bibfnamefont{J.}~\bibnamefont{Guan}},
  \bibinfo{author}{\bibfnamefont{K.}~\bibnamefont{Chen}},
  \bibinfo{author}{\bibfnamefont{S.}~\bibnamefont{C.Bae}}, \bibnamefont{and}
  \bibinfo{author}{\bibfnamefont{S.}~\bibnamefont{Granick}},
  \bibinfo{journal}{ACS Nano} \textbf{\bibinfo{volume}{7}},
  \bibinfo{pages}{9735} (\bibinfo{year}{2013}).

\bibitem[{\citenamefont{Bychuk and O'Shaughnessy}(1995)}]{Shaughnessy}
\bibinfo{author}{\bibfnamefont{O. V.}~\bibnamefont{Bychuk}} \bibnamefont{and}
  \bibinfo{author}{\bibfnamefont{B.}~\bibnamefont{O'Shaughnessy}},
  \bibinfo{journal}{Phys. Rev. Lett.} \textbf{\bibinfo{volume}{74}},
  \bibinfo{pages}{1795} (\bibinfo{year}{1995}).

\bibitem[{\citenamefont{Sokolov et~al.}(1997)\citenamefont{Sokolov, Mai, and
  Blumen}}]{Sokolov}
\bibinfo{author}{\bibfnamefont{I.~M.} \bibnamefont{Sokolov}},
  \bibinfo{author}{\bibfnamefont{J.}~\bibnamefont{Mai}}, \bibnamefont{and}
  \bibinfo{author}{\bibfnamefont{A.}~\bibnamefont{Blumen}},
  \bibinfo{journal}{Phys. Rev. Lett.} \textbf{\bibinfo{volume}{79}},
  \bibinfo{pages}{857} (\bibinfo{year}{1997}).

\bibitem[{\citenamefont{Lomholt et~al.}(2005)\citenamefont{Lomholt,
  Ambj\"{o}rnsson, and Metzler}}]{Lomholtpolymer}
\bibinfo{author}{\bibfnamefont{M.~A.} \bibnamefont{Lomholt}},
  \bibinfo{author}{\bibfnamefont{T.}~\bibnamefont{Ambj\"{o}rnsson}},
  \bibnamefont{and} \bibinfo{author}{\bibfnamefont{R.}~\bibnamefont{Metzler}},
  \bibinfo{journal}{Phys. Rev. Lett.} \textbf{\bibinfo{volume}{95}},
  \bibinfo{pages}{260603} (\bibinfo{year}{2005}).

\bibitem[{\citenamefont{Viswanathan et~al.}(1996)\citenamefont{Viswanathan,
  Afanasyev, Buldyrev, Murphy, Prince, and Stanley}}]{GMvish}
\bibinfo{author}{\bibfnamefont{G.~M.} \bibnamefont{Viswanathan}},
  \bibinfo{author}{\bibfnamefont{V.}~\bibnamefont{Afanasyev}},
  \bibinfo{author}{\bibfnamefont{S.~V.} \bibnamefont{Buldyrev}},
  \bibinfo{author}{\bibfnamefont{E.~J.} \bibnamefont{Murphy}},
  \bibinfo{author}{\bibfnamefont{P.~A.} \bibnamefont{Prince}},
  \bibnamefont{and} \bibinfo{author}{\bibfnamefont{H.~E.}
  \bibnamefont{Stanley}}, \bibinfo{journal}{Nature}
  \textbf{\bibinfo{volume}{381}}, \bibinfo{pages}{413} (\bibinfo{year}{1996}).

\bibitem[{\citenamefont{Viswanathan et~al.}(2011)\citenamefont{Viswanathan,
  Luz, Raposp, and Stanley}}]{GMVish_book}
\bibinfo{author}{\bibfnamefont{G.~M.} \bibnamefont{Viswanathan}},
  \bibinfo{author}{\bibfnamefont{M.~G. E.~D.} \bibnamefont{Luz}},
  \bibinfo{author}{\bibfnamefont{E.~P.} \bibnamefont{Raposp}},
  \bibnamefont{and} \bibinfo{author}{\bibfnamefont{H.~E.}
  \bibnamefont{Stanley}}, \emph{\bibinfo{title}{The Physics of Foraging: An
  Introduction to Random Searches and Biological Encounters}}
  (\bibinfo{publisher}{Cambridge University Press}, \bibinfo{year}{2011}).

\bibitem[{\citenamefont{Reynolds}(2009)}]{Reynolds1}
\bibinfo{author}{\bibfnamefont{A.~M.} \bibnamefont{Reynolds}},
  \bibinfo{journal}{J. Phys. A: Math. Theor.} \textbf{\bibinfo{volume}{42}},
  \bibinfo{pages}{434006} (\bibinfo{year}{2009}).

\bibitem[{\citenamefont{Costa et~al.}(2016)\citenamefont{Costa, Boccignone,
  Cauda, and Ferraro}}]{ferraro_PLOS}
\bibinfo{author}{\bibfnamefont{T.}~\bibnamefont{Costa}},
  \bibinfo{author}{\bibfnamefont{G.}~\bibnamefont{Boccignone}},
  \bibinfo{author}{\bibfnamefont{F.}~\bibnamefont{Cauda}}, \bibnamefont{and}
  \bibinfo{author}{\bibfnamefont{M.}~\bibnamefont{Ferraro}},
  \bibinfo{journal}{PLOS ONE} \textbf{\bibinfo{volume}{11}},
  \bibinfo{pages}{e0161702} (\bibinfo{year}{2016}).

\bibitem[{\citenamefont{Jespersen et~al.}(1999)\citenamefont{Jespersen,
  Metzler, and Fogedby}}]{Fogedby}
\bibinfo{author}{\bibfnamefont{S.}~\bibnamefont{Jespersen}},
  \bibinfo{author}{\bibfnamefont{R.}~\bibnamefont{Metzler}}, \bibnamefont{and}
  \bibinfo{author}{\bibfnamefont{H.~C.} \bibnamefont{Fogedby}},
  \bibinfo{journal}{Phys. Rev. E} \textbf{\bibinfo{volume}{59}},
  \bibinfo{pages}{2736} (\bibinfo{year}{1999}).

\bibitem[{\citenamefont{Mathai et~al.}(2010)\citenamefont{Mathai, Saxena, and
  Haubold}}]{Mathai}
\bibinfo{author}{\bibfnamefont{A.~M.} \bibnamefont{Mathai}},
  \bibinfo{author}{\bibfnamefont{R.~K.} \bibnamefont{Saxena}},
  \bibnamefont{and} \bibinfo{author}{\bibfnamefont{H.~J.}
  \bibnamefont{Haubold}}, \emph{\bibinfo{title}{{The H-Function, Theory and
  Applications}}} (\bibinfo{publisher}{Springer}, \bibinfo{year}{2010}).

\bibitem[{\citenamefont{Davies}(2002)}]{Gaver_stehfast}
\bibinfo{author}{\bibfnamefont{B.}~\bibnamefont{Davies}},
  \emph{\bibinfo{title}{Integral Transforms and Their Applications}}
  (\bibinfo{publisher}{Springer-Verlag NewYork Berlin Heidelberg},
  \bibinfo{year}{2002}).

\bibitem[{\citenamefont{Palyulin et~al.}(2013)\citenamefont{Palyulin, Chechkin,
  and Metzler}}]{search_levy_brown}
\bibinfo{author}{\bibfnamefont{V.~V.} \bibnamefont{Palyulin}},
  \bibinfo{author}{\bibfnamefont{A.~V.} \bibnamefont{Chechkin}},
  \bibnamefont{and} \bibinfo{author}{\bibfnamefont{R.}~\bibnamefont{Metzler}},
  \bibinfo{journal}{Proc. Natl. Acad. Sci. USA} \textbf{\bibinfo{volume}{111}},
  \bibinfo{pages}{2931} (\bibinfo{year}{2013}).

\bibitem[{\citenamefont{Hu et~al.}(2010)\citenamefont{Hu, Cheng, and
  Berne}}]{bruce_berne}
\bibinfo{author}{\bibfnamefont{Z.}~\bibnamefont{Hu}},
  \bibinfo{author}{\bibfnamefont{L.}~\bibnamefont{Cheng}}, \bibnamefont{and}
  \bibinfo{author}{\bibfnamefont{B.~J.} \bibnamefont{Berne}},
  \bibinfo{journal}{J. Phys. Chem.} \textbf{\bibinfo{volume}{133}},
  \bibinfo{pages}{034105} (\bibinfo{year}{2010}).

\bibitem[{\citenamefont{Janakiraman and
  Sebastian}(2014)}]{Deepika_sebastian_PRE2014}
\bibinfo{author}{\bibfnamefont{D.}~\bibnamefont{Janakiraman}} \bibnamefont{and}
  \bibinfo{author}{\bibfnamefont{K.~L.} \bibnamefont{Sebastian}},
  \bibinfo{journal}{Phys. Rev. E} \textbf{\bibinfo{volume}{90}},
  \bibinfo{pages}{040101(R)} (\bibinfo{year}{2014}).

\bibitem[{\citenamefont{Chechkin et~al.}(2007)\citenamefont{Chechkin,
  Sliusarenko, Metzler, and Klafter}}]{chechkin_numerical}
\bibinfo{author}{\bibfnamefont{A.~V.} \bibnamefont{Chechkin}},
  \bibinfo{author}{\bibfnamefont{O.~Y.} \bibnamefont{Sliusarenko}},
  \bibinfo{author}{\bibfnamefont{R.}~\bibnamefont{Metzler}}, \bibnamefont{and}
  \bibinfo{author}{\bibfnamefont{J.}~\bibnamefont{Klafter}},
  \bibinfo{journal}{Phys. Rev. E} \textbf{\bibinfo{volume}{75}},
  \bibinfo{pages}{041101} (\bibinfo{year}{2007}).

\end{thebibliography}
\end{document}